%% file: TVP-DR-Inflation.tex
\documentclass[12pt]{article}
\usepackage{amssymb}
\usepackage{amsthm} 
\usepackage{graphicx,psfrag,epsf,epstopdf}
\usepackage{enumerate}
\usepackage{siunitx}
\usepackage{natbib} 
\usepackage{amsmath,amsfonts,amssymb,amsthm,algorithm}
\usepackage{bm}    
\usepackage{threeparttable}   
\usepackage{tcolorbox}    
\usepackage{color,float}    
\usepackage{multirow}    
\usepackage{caption}
\usepackage{subcaption}
\usepackage[toc,page]{appendix}    
\usepackage{hyperref}
\hypersetup{     
	colorlinks=true, 
	linkcolor=blue,
	citecolor=blue 
}
\usepackage{setspace}

\usepackage[figuresright]{rotating}

\usepackage{adjustbox}
\definecolor{darkblue}{RGB}{0,0,139}

\newtheorem{remark}{Remark}  

\newcounter{example counter}

\def \R {\mathbb{R}}

\def\1{1\!{\rm l}}
\DeclareMathOperator{\diag}{diag}
\DeclareMathOperator{\vecop}{vec}
\definecolor{db}{rgb}{0, 0, 0.55}

\author{
	Yunyun Wang\footnote{
		College of Finance and Statistics, Hunan University    
		(\href{mailto:yunyunwang@hnu.edu.cn}{yunyunwang@hnu.edu.cn}).} 
	\and
	Tatsushi Oka\footnote{
		Department of Economics, Keio University    
		(\href{mailto:tatsushi.oka@monash.edu}{tatsushi.oka@keio.jp}).} 
	\and 
	Dan Zhu\footnote{Department of Econometrics and Business Statistics, Monash University    
		(\href{mailto:dan.zhu@monash.edu}{dan.zhu@monash.edu}). Corresponding Author.} 
	\vspace{0.5cm}
}

\begin{document}
	\title{Inflation Target at Risk: A Time-varying Parameter Distributional Regression\thanks{We are grateful to Christiane Baumeister, Joshua Chan, Jamie Cross, Jiti Gao, Chenhan Hou, Dimitris Korobilis, Gary Koop, James Morley and Barbara Rossi for their valuable comments and suggestions. We also wish to acknowledge the insightful discussions and feedback received at the 17th International Conference on Computational and Financial Econometrics (CFE 2023),  Annual Conference of the International Association for Applied Econometrics (IAAE 2024, Xiamen),  Econometric Society Australasia Meeting (ESAM 2024), and the 8th International Conference on Econometrics and Statistics (EcoSta 2025). Oka gratefully acknowledges financial support by the Japan Society of the Promotion of Science under KAKENHI 23K18807.}}
	\maketitle

	\begin{abstract}	
    Inflation exhibits state-dependent, skewed, and fat-tailed dynamics that make risk a central concern for monetary policy. Accordingly, inflation risks are distributional and cannot be fully captured by mean-based models. We propose a flexible time-varying parameter distributional regression model that estimates the full conditional distribution of inflation, allowing macroeconomic drivers to have nonlinear and asymmetric effects across the distribution. Applied to U.S. inflation, the model captures major shifts in tail-risk probabilities. Analysis of risk drivers shows that deflationary pressures arise primarily from demand-side weakness and inflation persistence, whereas upside risks are driven mainly by supply-side shocks, particularly energy price inflation. Examining the impact of key drivers further reveals that the unemployment-inflation relationship weakens in the distributional tails. Energy price shocks, by contrast, have little effect on deflation risk but exhibit strongly time-varying and asymmetric effects on high-inflation risk.
	\end{abstract}
	
	\vspace{0.5cm}
	\noindent%
	 \vspace{0.2cm} 
	
	\setstretch{1.3}

\thispagestyle{empty}
\newpage
\setcounter{page}{1}
\section{Introduction}
  Uncertainty is the defining feature of the monetary policy landscape, arising from a wide range of possible macroeconomic outcomes and the continual evolution of economic structure. Risk management, therefore, lies at the heart of effective policymaking, particularly for monetary policy \citep{greenspan2004risk}. This imperative is evident in the case of inflation. The prolonged low-inflation period following the Global Financial Crisis and the abrupt inflationary surge after the pandemic both revealed large and persistent deviations from the expected paths. Moreover, the underlying distribution exhibited strong skewness and fat tails during these episodes \citep{harding2022resolving,lopez2024inflation}. These patterns highlight the state-dependent nature of inflation dynamics and underscore the importance of time-varying tail risks for monetary policy decisions.

 Building on these insights, this paper proposes a framework for assessing inflation target risks, defined as the risks that future inflation deviates from the central bank's long-run target. \citet{kilian2007quantifying} articulate that inflation target risks are inherently embedded in the full conditional distribution rather than the conditional mean alone. Meanwhile, time-varying parameter models with stochastic volatility \citep{primiceri2005time,stock2007has,chan2017stochastic}  demonstrate the importance of allowing means and variances to evolve over time. These models often impose Gaussian and symmetric errors, limiting variation in skewness and tail thickness relevant to inflation risk assessment. Our framework bridges these two literature through a time-varying parameter distributional regression (TVPDR) model that directly estimates the entire conditional distribution of inflation given macroeconomic variables, allowing their effects to vary nonlinearly and asymmetrically across the distribution.

 We analyze U.S. inflation under the TVPDR framework using Consumer Price Index (CPI) data from 1982:Q1 to 2024:Q4, conditioning on a broad set of macroeconomic predictors. Our results reveal pronounced shifts in inflation target risks over time: deflation risk rose sharply during the Great Recession and pandemic, while excessive inflation risk surged post-pandemic. Decomposing these risks by their drivers, we find that
 excessive inflation risk originates predominantly from supply-side disturbances, especially energy and food price shocks, whereas deflationary risk arises almost exclusively from adverse demand conditions. These findings complement recent work on the sources of inflation risk: commodity price movements \citep{garratt2022commodity}, unexpectedly strong post-pandemic demand \citep{giannone2024drivers}, and non-wage supply shocks \citep{bernanke2025caused}. Our analysis documents the propagation of these forces throughout the inflation distribution and their role in shaping distinct inflation-target risks, thereby reconciling prior empirical findings.

We also offer a distributional perspective on Phillips curve (PC) flattening and state dependence.The literature documents a pronounced flattening of the PC slope from the high-inflation 1970s–1980s to the low-inflation 2000s and 2010s \citep[e.g.][]{blanchard2016phillips, delnegro2020whats, barnichon2020identifying}. Our results suggest that changes in unemployment have negligible effects on  high-inflation risk and only weak effects on deflation risk, mainly before 2021. This suggests that PC flattening is a distributional phenomenon rather than one confined to the mean. Cost-push forces, in contrast, exhibit distinct distributional effects. Energy price variation has little impact on deflation risk but has strong, time-varying, and asymmetric effects on high-inflation risk. These findings align with recent evidence that inflation responsiveness to slack and supply shocks depends on the prevailing inflationary regime \citep{forbes2021low, harding2023}.

  The proposed method is well suited for risk assessment under target-based policy. Since central banks set inflation targets as specific numerical values or ranges, such as 2\%, our approach directly estimates the conditional distribution to assess these target risks. A strand of recent literature on quantile regression \citep[e.g.][]{korobilis2021time,lopez2024inflation}, in contrast, focuses on selected tail quantiles. These methods require subsequent transformation, typically parametric approximations, to recover a full predictive density from which target risks are computed. Our approach complements this literature by directly estimating the full conditional distribution, avoiding additional approximation in line with Vapnik's dictum to ``solve the problem of interest'' directly \citep{Vapnik1998}. We provide empirical support for this approach through a forecasting comparison against a comprehensive set of benchmarks, where the TVPDR model delivers superior accuracy for distributional features.

 The remainder of this paper proceeds as follows. After reviewing related literature in Section \ref{sec: Motivation}, we present the model and risk measures in Section \ref{sec: Model}. Section \ref{sec: Application} reports forecasting performance and risk analysis for U.S. inflation. Section \ref{sec: Con} concludes. The appendices contain additional details on estimation and data.

\section{Related Literature}\label{sec: Motivation}
This paper contributes to two primary strands of research: the evolving literature on the PC and the growing body of work on inflation risk analysis. We discuss the related literature in this section.

\subsection{Time-varying and Nonlinear Phillips Curve}
  The PC provides a framework for linking inflation to real economic conditions, particularly labor market slack, and is widely used to analyze inflation dynamics and forecast inflationary pressures. A long-standing and central challenge in empirical macroeconomics is the profound instability of the PC slope, which governs the short-run sensitivity of inflation to economic slack and thus directly affects assessments of inflation risks and the efficacy of monetary policy.

    \begin{table}[H]
    	\centering
    	\setlength{\tabcolsep}{4pt} 
    	\renewcommand{\arraystretch}{1.2} 
    	\caption{Selected Empirical Estimates of the Phillips Curve Slope}
    	\label{tab:PC_estimates}
    	\resizebox{\textwidth}{!}{ 
    	\begin{tabular}{llll}
    		\hline
    		\textbf{Study}                                   & \textbf{Variable (sample)}        & \textbf{Method (Simplified)} & \textbf{PC Slope Estimates} \\ \hline
    		\cite{ball2019phillips}         & Median of industry inflation      & Expectations-augmented     & -0.76                       \\
    		& (1985-2015)                       &              PC                &                             \\ 
    		[2pt] 
    		\cite{barnichon2020identifying} & Core PCE (1969-2017)              & Hybrid NKPC using IV         & -0.45 (1969-2007);          \\
    		&                                   &                              & -0.24 (1990-2017)           \\ 
    		[2pt] 
    		\cite{blanchard2016phillips}    & Headline CPI (1960-2014)          & TVP PC                       & -0.7 (1970s);               \\
    		&                                   &                              & -0.2 (post-1980s)           \\
    		[2pt] 
    		\cite{coibion2015phillips}      & CPI/GDP Deflator       & Standard PC                  & -0.517 for CPI;             \\
    		&                  (1960-2007)                 &                              & -0.399 for GDP Deflator     \\
    		[2pt] 
    		\cite{cerrato2022inflation}     & Core CPI (1990-2022)              & Two-region NKPC              & -0.25 (Pre-COVID);          \\
    		&                                   &                              & 0.02 (COVID);               \\
    		&                                   &                              & -0.85 (Post-COVID)          \\
    		[2pt] 
    		\cite{cristini2021nonlinear}    & CPI (1961-2019)                   & Piecewise model              & -0.32 (tight market);       \\
    		&                                   &                              & -0.06 (slack market)        \\
    		&                                   & Threshold model              & -0.6 (tight market);        \\
    		&                                   &                              & -0.1 (slack market)         \\
    		[2pt] 
    		\cite{crump2024unemployment}    & Price \& Wage Inflation           & Micro–macro PC               & Median of 0.03;             \\
    		& (1960-2019)                       &                              & Range of 0.02–0.06          \\
    		[2pt] 
    		\cite{delnegro2020whats}        & Core PCE (1973-2019)              & SVAR \& DSGE                 & -0.3 (1973-1989);           \\
    		&                                   &                              & Near 0 (1989-2019)          \\
    		[2pt] 
    		\cite{doser2023inflation}       & CPI (1968-2019)                   & Threshold model              & -0.5 (tight market);        \\
    		&                                   &                              & 0.08 (slack market)         \\
    		[2pt] 
    		\cite{fitzgerald2024stable}     & Regional Headline CPI  & New Keynesian                & 0.276 (state-level);        \\
    		&              (1977-2018)                     &                              & 0.008 (aggregate)           \\
    		[2pt] 
    		\cite{gordon2013phillips}       & PCE (1962-2013)                   & Triangle model/ NKPC         & -0.5 (Triangle);            \\
    		&                                   &                              & -0.2 (NKPC)                 \\
    		[2pt] 
    		\cite{hazell2022slope}          & Nontradeable Price Inflation      & Regional (state-level) PC    & 0.0062                      \\
    		& (1978-2018)                       &                              &                             \\
    		[2pt] 
    		\cite{inoue2025flattened}       & GDP Deflator (1970-2021)          & TVP PC using IV              & Decreased since 1980s;      \\
    		&                                   &                              & Near zero (2010-2020);      \\
    		&                                   &                              & increase to 0.009 (2024)    \\ [2pt]    		
    		\cite{smith2025breaks}          & CPI (1980-2022)                   & Bayesian PC                  & A kink in PC;               \\
    		&                                   &                              & -0.26 (before 2000);        \\
    		&                                   &                              & -0.21 (after 2000)          \\ \hline
    		\end{tabular}}
    		\begin{minipage}{1\linewidth} 
    			\linespread{1}\footnotesize
    			\textit{Notes}: Forcing variables considered in these studies are the unemployment gap.
    		\end{minipage}
    \end{table}    
   
   Table \ref{tab:PC_estimates} summarizes representative empirical estimates of the PC slope across studies, with a focus on the U.S. inflation.  The results show that slope estimates differ not only across sample periods and inflation measures (headline vs. core) but also across model specifications, constant-parameter, TVP, or nonlinear forms. A primary finding in the literature is the structural flattening of the curve since the 1980s. \cite{blanchard2016phillips} and \cite{delnegro2020whats} document a sharp decline in the sensitivity of inflation to unemployment, a trend often attributed to more effectively anchored inflation expectations and increased central bank credibility \citep{ball2019phillips,coibion2015phillips}. However, others argue that this flattening may be an artifact of measurement and identification issues. \cite{gordon2013phillips} and \cite{barnichon2020identifying} find that when accounting for supply shocks or using structural monetary shocks as instruments, the underlying slope appears much steeper and more stable than standard estimates suggest. This view is further supported by regional data, which reveals that state-level PC are significantly steeper than national aggregates \citep{hazell2022slope,fitzgerald2024stable}. 
   
   A complementary strand of literature emphasizes that the PC is state-dependent and nonlinear. \cite{cristini2021nonlinear} and \cite{smith2025breaks} find that the curve is often kinked or convex, meaning inflation is far more responsive in tight labor markets than during periods of economic slack, while \cite{doser2023inflation} suggest that once inflation expectations are measured so as to account for consumer expectations, nonlinearities in PC are muted overall. On the other hand, \cite{cerrato2022inflation} and \cite{inoue2025flattened} document a dramatic post-pandemic re-steepening of the curve, while \cite{crump2024unemployment} argue this surge was driven by shifts in the natural rate of unemployment rather than the slope itself. Despite these advances, the functional form of the PC remains sensitive to modeling choices and continues to be the subject of active empirical debate.
   
  \subsection{From Mean to Distributional Dynamics}
   From a risk-assessment perspective, the critical decisions for policymakers, such as timely tightening or easing, hinge on quantifying and managing tail risks, namely the probabilities of deflation (left-tail risk) and excessively high inflation (right-tail risk). These policy-relevant risks are embedded entirely within the full conditional distribution.
   
   The literature documented that looking at the entire predictive distribution of inflation can reveal additional insights into their dynamics and risk analysis.  \cite{kilian2007quantifying} propose formal and quantitative measures of the risk that future inflation will be excessively high or low relative to the range preferred by a private sector agent, which is defined based on the distribution function of inflation. \citet{kilian2008central} show that U.S. monetary policy under Chairman Greenspan is better characterized by the Federal Reserve’s systematic weighting of upside and downside inflation risks, rather than by responses to the conditional mean of inflation and the output gap alone. \cite{andrade2012tails} introduce a new measure of inflation risks derived from survey-based density forecasts, showing that the asymmetry and magnitude of inflation risks evolve over time and significantly impact future inflation and central bank interest rate targets.
   	
   	In recent years, QR models are widely adopted to study the inflation tail risks. For example, \cite{korobilis2017quantile} shows the efficacy of combining a set of Bayesian QR estimates for inflation forecasting. \cite{korobilis2021time} and \cite{pfarrhofer2022modeling} explore the tail risks of inflation based on TVP-QR models within a Bayesian framework. More recently, based on univariate and panel QR models, \cite{lopez2024inflation} show that fluctuations in this risk are closely linked to variations in the intensity of cost-push shocks, which can also have asymmetric effects across the distribution of possible inflation outcomes. While QR models address distributional asymmetries, they generally focus on specific tail quantiles and require a second-stage procedure, such as fitting a parametric distribution, to recover a full predictive density.

   	Recent research also provides strong evidence that the responsiveness of inflation to economic slack, expectations, and cost-push shocks varies substantially across inflationary environments. For example, \cite{gagnon2019low} suggest that the PC bends so that excessively high unemployment has less effect on inflation than excessively low unemployment, and this bend only becomes apparent when inflation is very low. \cite{forbes2021low} show that the PC is normally steep but becomes significantly flat only when inflation is very low and economic slack is high. Extending this perspective, \cite{harding2023} propose a nonlinear PC that has a flat slope when inflationary pressures are subdued and steepens when inflationary pressures are elevated. They find a stronger transmission of shocks when inflation is high, which generates conditional heteroskedasticity in inflation and inflation risk. These findings suggest that the essential asymmetric and state-dependent nature of inflation dynamics emerges across the full spectrum of outcomes, not only the central tendency.
   	
   We propose a TVPDR framework that directly models the full conditional distribution of inflation without restrictive parametric assumptions. By allowing macroeconomic drivers to exert nonlinear and asymmetric effects across the distribution, our approach provides a unified framework for analyzing inflation dynamics and quantifying policy-relevant inflation risks.

  \section{Time-varying Parameter Distributional Regression} \label{sec: Model}
    
   In this section, we introduce our TVPDR model and inflation risk measures. The TVPDR framework is specifically tailored to analyze distribution dynamics by combining the flexibility of time-varying parameters with a distributional regression (DR) structure \citep{foresi1995conditional}. 
  	
    The key advantage of DR is its ability to model the full conditional distribution directly in a data-driven manner, without imposing restrictive parametric assumptions. Building on recent work by \citet{wang2023distributional}, who extend DR to time-series settings by modeling multivariate conditional distributions for stationary processes, this paper further advances the approach by allowing regression parameters to evolve over time. This extension enables the model to capture structural instabilities as well as time variation in the shape of the conditional distribution, including changes in skewness, tail behavior, and other higher-order features.

\subsection{Model Specifications}

In the following, $\1\{\cdot\}$ denotes the indicator function, which equals 1 if the condition inside $\{\cdot\}$ is satisfied and 0 otherwise. We use $\mathbf{0}_n$ to represent an $n \times 1$ zero vector, $\mathbb{O}_n$ for an $n \times n$ zero matrix, $\mathbb{I}_n$ for the $n \times n$ identity matrix, and $\mathbb{I}_{n, -1}$ for an $n \times n$ matrix with ones on the second subdiagonal and zeros elsewhere. The symbol ``$\otimes$'' denotes the Kronecker product, while ``$\preceq$'' and ``$\succeq$'' indicate element-wise inequality between vectors. The operator $\vecop(\cdot)$ stacks the columns of a matrix into a vector, and $\diag(\cdot)$ either forms a diagonal matrix from a vector or a block-diagonal matrix from a list of matrices.

Let $Y_t$ be a time-series variable with support $\mathcal{Y}_t$ and $X_t$ be a $k\times 1$ vector of appropriately lagged predictors with support $\mathcal{X}_t$. We denote the conditional distribution function of $Y_t$ given $X_t$ as $F_{Y_t|X_t}$. The proposed TVPDR model characterizes this conditional distribution by modeling it at an arbitrary value $y\in\mathcal{Y}_t$ as follows,
\begin{align}\label{TVPDR}
	F_{Y_t|X_t}(y|X_t)=\Lambda\left(g(X_t)^\top\beta_{y, t}\right)
\end{align}
where $\Lambda: \R \to [0,1]$ is a known link function such as logit, probit, and log-log, $g: \mathbb{R}^{k}\rightarrow \mathbb{R}^{d}$ is a known transformation of the conditioning variables such as polynomials, b-splines, and tensor products, and $\beta_{y, t}=(\beta_{y,t,1},\ldots, \beta_{y,t,d})^\top$ is a $d\times 1$ vector of time-varying parameters specific to threshold $y$.\footnote{As shown in \cite{chernozhukov2013inference}, for a sufficiently rich transformation of the covariates, one can approximate the conditional distribution function arbitrarily well without extra concern about the choice of the link function. The decision to apply transformations to regressors in a regression model should be guided by statistical assumptions and practical aspects of model interpretation and complexity.} In particular, when the parameters $\beta_{y,t}$ become constant over time, the model simplifies to the standard DR model, which can be estimated as a binary choice model for the binary outcome $\1\{Y_t \le y\}$ under the maximum likelihood framework \citep{chernozhukov2013inference}.

We employ the random walk assumption as a foundational component of our TVP model. That is, we consider $\beta_{y, t}$ evolves according to a random walk with Gaussian error: 
\begin{align}\label{eq: TVP}
	\beta_{y, t} =\beta_{y,t-1}+\eta_{y, t},\ \ \eta_{y, t}\sim \mathcal{N}(\mathbf{0}_{d},\Sigma_{y}),
\end{align}
where the process is initialized with $\beta_{y,1}\sim \mathcal{N}(\mathbf{0}_{d},\Sigma_{y})$, and $\Sigma_{y}$ is a $d \times d$ covariance matrix that governs the time-variation of $ \beta_{y, t}$. 
Using random walk evolution in TVP models is a popular approach in econometrics and finance for capturing the dynamics of parameters that change over time \citep{cogley2005drifts,primiceri2005time,nakajima2011time,inoue2025flattened}. It is useful for capturing the permanent shifts in the parameters, such as long-term trends or structural changes, and can reduce the complexity of the estimation procedure.

If the object of interest is restricted to specific probabilities, such as the likelihood that inflation falls below or exceeds a given threshold, then applying one or two targeted TVPDR regressions is sufficient. However, to recover the entire conditional distribution function, $F_{Y_t|X_t}(\cdot|X_{t})$, it is necessary to apply the TVPDR to model this function across a sequence of discrete points that are sufficiently fine over the support $\mathcal{Y}_t$. The resulting collection of estimation results constitutes an approximation of the entire conditional distribution of $Y_t$. 

\begin{remark}
	The proposed framework specifies a time-varying conditional cumulative distribution function ($F_{Y_t|X_t}$) for the outcome variable, which naturally and simultaneously accommodates dynamic variation in all conditional moments over time. For instance, the conditional mean and variance are derived by integrating over $F_{Y_t|X_t}$:
	\begin{align*}
		\mathbb{E}\left(Y_t|X_t\right)=\int y d F_{Y_t|X_t}(y|X_t),\ \ \ \ 
		Var\left(Y_t|X_t\right)=\int y^2 d F_{Y_t|X_t}(y|X_t)-[\mathbb{E}\left(Y_t|X_t\right)]^2.
	\end{align*}	
	Consequently, since the distribution $F_{Y_t|X_t}$ is governed by the time-varying coefficients $\beta_{y,t}$, the derived moments will dynamically evolve as functions of the estimated parameters and the conditioning variables.
\end{remark}

This full distributional approach, however, necessitates estimating a substantial proliferation of time-varying parameters. Furthermore, to guarantee that the approximated function is a statistically valid cumulative distribution function, we must enforce the essential monotonicity constraint that $F_{Y_t|X_t}$ does not decrease as $y$ increases. We introduce an efficient Markov Chain Monte Carlo (MCMC) sampler to simultaneously manage the computational load and enforce this constraint. The full details regarding this estimation framework and the MCMC algorithm are provided in Appendix \ref{sec: Estimation}.

\subsection{Inflation Target and Risk Measures} \label{subsec: risk measure}

 Modern central banks increasingly operate under a risk-management paradigm in which policy decisions are guided by the entire distribution of future inflation outcomes rather than by point forecasts alone. This shift is reflected in the routine use of distribution forecasts, often communicated through fan charts, that explicitly quantify uncertainty, asymmetry, and tail risks surrounding macroeconomic projections. 
 
 In the U.S., for example, the FOMC now frames its inflation outlook in explicitly probabilistic terms. The Summary of Economic Projections (SEP) reports histograms of individual forecasts and fan charts displaying 70\% confidence intervals around the median inflation path,\footnote{\href{https://www.federalreserve.gov/monetarypolicy/files/fomcprojtabl20250917.pdf}{Summary of Economic Projections – September 2025. Board of Governors of the Federal Reserve System.}} while recent SEPs and FOMC Minutes regularly emphasize the balance of upside and downside risks to inflation.\footnote{\href{https://www.federalreserve.gov/monetarypolicy/fomcminutes20250917.htm}{Minutes of the Federal Open Market Committee – September 2025. Board of Governors of the Federal Reserve System.}} In practice, central banks seek to stabilize inflation around a long-run objective, commonly 2\% in the U.S., with growing recognition that policy is effectively conducted with respect to a range rather than a point target \citep{mankiw2024six}. Deviations from the target can carry substantial macroeconomic costs: risks of persistent undershooting can weaken demand and destabilize expectations, while risks of sustained overshooting can erode purchasing power and complicate policy normalization. 
 
 To quantify these risks formally in a distributional framework, we follow the approach of \citet{kilian2007quantifying}, which characterizes inflation risks by jointly accounting for both the probability and the magnitude of deviations from a policy-relevant target range. Specifically, let $F_{\pi_{t}|X_t}$ denote the conditional distribution of inflation given the state of the economy at time $t$, $[\underline{\pi}, \bar{\pi}]$ be the preferred range of inflation, where $\underline{\pi} < \bar{\pi}$ are fixed inflation thresholds. The deflation and excessive inflation risk of $\pi_{t}$ are defined as
\begin{align}
    & DR_{t}(\underline{\pi},\alpha) := - \int_{-\infty}^{\underline{\pi}} (\underline{\pi} - \pi)^\alpha \, d F_{\pi_{t}|X_t}(\pi \mid x_t), \\
    & EIR_{t}(\bar{\pi},\gamma) := \int_{\bar{\pi}}^{\infty} (\pi - \bar{\pi})^\gamma \, d F_{\pi_{t}|X_t}(\pi \mid x_t),
\end{align}
where the parameters $\alpha \geq 0$ and $\gamma \geq 0$ capture the degree of risk aversion of the economic agent.  These metrics generalize familiar tail-risk concepts. When $\alpha=\gamma=0$, they reduce to simple probabilities of falling below $\underline{\pi}$ or exceeding $\bar{\pi}$. For $\alpha,\gamma > 0$, they weight more extreme deviations more heavily, allowing analysts to emphasize severe tail events. For example, when $\alpha=\gamma=1$, they can be interpreted as measures of expected deflation, $DR_t(\underline{\pi},1)=E(\pi_{t}-\underline{\pi}\mid\pi_{t}<\underline{\pi},x_t)P(\pi_{t}<\underline{\pi}\mid x_t)$, and expected excessive inflation $EIR_t(\underline{\pi},1)=E(\pi_{t}-\overline{\pi}\mid\pi_{t}>\overline{\pi},x_t)P(\pi_{t}>\overline{\pi}\mid x_t)$. This flexibility allows policymakers and analysts to tailor the risk metrics to different tolerance levels for inflation instability, providing a unified framework for assessing the distributional implications of macroeconomic shocks.

It is worth noting that both measures are defined as functionals of the conditional distribution. Consequently, their accuracy depends critically on how well the model captures the entire shape of the distribution, including its tails, asymmetry, and time variation. While recent studies employing QR have explored such risk measures \citep{korobilis2021time, lopez2024inflation}, these approaches typically require a second-stage parametric approximation, such as fitting a skewed-t distribution to discrete quantiles, to recover a continuous distribution. In contrast, the TVPDR framework models the conditional distribution directly in a data-driven manner, ensuring internal consistency across probabilities and tail risks while allowing the full shape of the distribution to evolve over time. This makes TVPDR particularly well-suited for distribution-based inflation risk measurement and policy analysis.

\section{U.S. Inflation Forecasting and Risks Analysis}\label{sec: Application}
In this section, we apply the TVPDR approach to explore the conditional distribution of inflation, assess the risks of future inflation deviating significantly above or below the target range, and examine the influence of macroeconomic drivers on these inflation risks.

\subsection{Data and Model Specifications}
We analyze both headline and core inflation, measured using the U.S. Consumer Price Index (CPI), with quarterly observations spanning from 1982:Q1 to 2024:Q4. Model estimation and evaluation are based on an expanding window approach, with the initial estimation period covering 1982:Q1 to 1999:Q4. Let $P_t$ be the quarterly CPI at time $t$, the $h$-quarter-ahead annualized inflation rate, $\pi_{t+h}:=(400/h)\ln(P_{t+h}/P_{t})$, is used in our analysis as the dependent variable in the $h$-step-ahead inflation forecasting model.

We begin by introducing our suite of benchmark models, which encompasses both traditional mean-based regression approaches and more flexible quantile regression frameworks, featuring constant or time-varying parameters. The models and their specifications are detailed below:

\begin{enumerate}
	\item  Linear regression (LR): A linear regression model with constant parameters and homoskedastic Gaussian errors: $\pi_{t+h}=X_t^\top\beta+e,\ e\sim N(0, \sigma^2)$.
	\item Quantile regression (QR): A quantile regression model with constant parameters: $Q_\tau(\pi_{t+h}|X_t)=X_t^\top\beta_\tau $.
	\item Distributional regression (DR): A distributional regression model with constant parameters: $F_{\pi_{t+h}}(y|X_t)=\Phi(X_t^\top\beta_y)$.
	\item Unobserved component stochastic volatility (UCSV): A model where inflation ($\pi_{t+h}$) is decomposed into an unobserved, slowly moving trend ($\tau_t$) and a transitory component, with the variance of both components potentially following a stochastic volatility process: $\pi_{t+h} = \tau_{t} + \varepsilon_{t}, \quad \varepsilon_{t} \sim N(0, \exp(h_t))$; $\tau_{t} = \tau_{t-1} + \eta_{t}, \quad \eta_{t} \sim N(0, \exp(g_t))$.
	\item  Time-varying parameter stochastic volatility (TVPSV): A linear regression model with time-varying parameters and stochastic volatility, where both the parameters and log-variances follow random walk processes: $\pi_{t+h}=X_t^\top\beta_t+e_t,\ \ e\sim N(0, \exp(h_t))$, where $\beta_{t}=\beta_{t-1}+u_t,\ u_t\sim N(0, \Omega);\ \ h_t=h_{t-1}+\eta_t,\ \eta_t\sim N(0, \sigma^2_\eta)$.
	\item  Time-varying parameter quantile regression (TVPQR): A quantile regression model with time-varying parameters evolving as random walks: $Q_\tau(\pi_{t+h}^h|X_t)=X_t^\top\beta_{t,\tau} $, where $\beta_{t,\tau}=\beta_{t-1,\tau}+u_{t,\tau},\ u_{t,\tau}\sim N(0, \Sigma_{\tau})$.
\end{enumerate}

 All benchmark models, along with the TVPDR model, incorporate the same set of covariates: a constant term, three PC determinants (one-quarter-lagged inflation, five-year inflation expectations, and the unemployment rate), as well as energy price inflation, food price inflation, national financial condition index, Federal funds rate, and real GDP growth. Detailed descriptions and data resources of each time series variable are provided in Appendix \ref{sec: Appendix}.

 The LR model is estimated by ordinary least squares. DR and QR models with constant parameters are estimated using maximum likelihood estimation and linear programming, respectively. All time-varying or stochastic volatility models are estimated using Bayesian MCMC methods with 10,000 posterior draws following a 5,000 iteration burn-in period, with convergence assessed via standard diagnostics. For the distributional regression approaches (DR and TVPDR), we estimate conditional probabilities across a fine grid of thresholds spaced at 0.1\% intervals spanning the full range of observed inflation values, while the quantile-based models (QR and TVPQR) are estimated at all percentiles from $\tau = 0.01$ to $\tau = 0.99$ in 0.01 increments. 
	
	\subsection{Out-of-sample Performance}
	This subsection assesses the out-of-sample forecasting performance of each model over forecast horizons $h = 1$ to $4$, using a range of metrics that capture both point and distributional aspects of forecast performance.
	
    For point forecast evaluation, we employ the Root Mean Squared Error (RMSE):
    {\small
	\[
	\text{RMSE}_{h}=\sqrt{\frac{1}{T}\sum_{t=1}^T\left(\pi_{t+h}-E(\pi_{t+h} |X_t)\right)^2}.
	\]}
    To evaluate the full forecasting distributions, we compute the Continuous Ranked Probability Score (CRPS):
    {\small
	\begin{align*}
		\text{CRPS}_{h}= \frac{1}{T}\sum_{t=1}^{T}\int_{-\infty}^{\infty} \left( \widehat{F}_{\pi_{t+h}|X_t}(\pi|x_t) - \1\{\tilde{\pi}_{t+h} \leq \pi\} \right)^2 d\pi,
	\end{align*}}
   where $\widehat{F}_{\pi_{t+h}|X_t}(\pi|x_t)$ is the estimated conditional distribution function of $\pi_{t+h}$ at location $\pi$, and $\tilde{\pi}_{t+h}$ is the true realization of $\pi_{t+h}$. Tails of the inflation distribution are very important for studying the inflation risks. We use the quantile score (QS) as a measure to assess the tail forecasting accuracy, which is defined by,
	{\small
	\[
	QS_{h}(\tau) :=\frac{1}{T}\sum_{t=1}^T\big[\tilde{\pi}_{t+h}-\widehat{Q}_{\tau}(\pi_{t+h}|x_t)\big]\1 \big\{\tilde{\pi}_{t+h}\leq \widehat{Q}_{\tau}(\pi_{t+h}|x_t)\big\},  
	\]}
	where $\widehat{Q}_{\tau}(\pi_{t+h}|x_t)$ is the estimated $\tau$-th conditional quantile of $\pi_{t+h}$. Smaller values of the loss function indicate better performance. We compare the QS for both the 5\% and 95\% quantiles. 
	
	The Great Recession (2008-2009) and pandemic recession (2020-2022) produced extreme realizations in both headline and core inflation, where conventional forecasting models systematically underperform, generating forecast errors several times larger than in stable periods. Considering the substantial impact of these errors on the overall performance, in the following comparison, we present separate results for: (1) Recession Periods (2008:Q1-2010:Q4 and 2020:Q1-2022:Q4) and (2) Non-Recession Periods (all other quarters in our sample). This disaggregation allows for clearer evaluation of model performance across different economic regimes.

	\begin{table}[H]
		\footnotesize
		\centering
		\caption{Forecasting Performance Comparison for Headline CPI Inflation}
		\begin{tabular}{lllllllllll}
			\hline
			&       & \multicolumn{4}{c}{Non-Recession Periods}                         &  & \multicolumn{4}{c}{Recession Periods}                             \\ \cline{3-6} \cline{8-11} 
			&       & $h=1$          & $h=2$          & $h=3$          & $h=4$          &  & $h=1$          & $h=2$          & $h=3$          & $h=4$          \\ \hline
			\textit{RMSE}     & LR    & 1.620          & 1.619          & 1.622          & 1.621          &  & 3.689          & 3.951          & 4.051          & 3.994          \\
			& QR    & 1.796          & 1.815          & 1.823          & 1.769          &  & 4.961          & 4.756          & 5.079          & 4.639          \\
			& DR    & 1.731          & 1.619          & 1.616          & 1.666          &  & \textbf{3.471} & \textbf{3.509} & 3.948          & 4.276          \\
			& UCSV  & 1.582          & 1.585          & 1.606          & \textbf{1.592} &  & 3.744          & 3.650          & \textbf{3.658} & \textbf{3.734} \\
			& TVPSV & 1.770          & 1.690          & 1.666          & 1.698          &  & 5.006          & 4.757          & 4.028          & 4.480          \\
			& TVPQR & 1.592          & 1.602          & 1.628          & 1.609          &  & 4.471          & 4.146          & 4.013          & 4.684          \\
			& TVPDR & \textbf{1.534} & \textbf{1.568} & \textbf{1.538} & 1.609          &  & 3.705          & 3.709          & 3.896          & 3.961          \\ \hline
			\textit{CRPS}     & LR    & 0.905          & 0.900          & 0.899          & \textbf{0.902} &  & 1.975          & 2.234          & 2.290          & 2.569          \\
			& QR    & 1.140          & 1.136          & 1.210          & 1.183          &  & 2.073          & 2.247          & 2.276          & 2.600          \\
			& DR    & 0.992          & 0.928          & 0.919          & 0.956          &  & 2.572          & 2.942          & 2.493          & 2.762          \\
			& UCSV  & 0.910          & 0.906          & 0.916          & 0.912          &  & 1.917          & \textbf{1.896} & \textbf{1.912} & \textbf{1.938} \\
			& TVPSV & 1.825          & 1.890          & 1.903          & 1.860          &  & 3.295          & 3.337          & 3.426          & 3.472          \\
			& TVPQR & 0.926          & 0.927          & 0.945          & 0.922          &  & 2.195          & 2.096          & 2.253          & 2.181          \\
			& TVPDR & \textbf{0.883} & \textbf{0.888} & \textbf{0.881} & 0.923          &  & \textbf{1.910} & 1.980          & 2.147          & 2.135          \\ \hline
			\textit{QS: 5\%}  & LR    & \textbf{0.219} & 0.233          & 0.265          & 0.273          &  & 0.663          & 0.699          & 0.728          & 0.683          \\
			& QR    & 0.285          & 0.291          & 0.285          & 0.278          &  & 0.814          & 0.815          & 0.943          & 0.804          \\
			& DR    & 0.275          & 0.282          & 0.251          & 0.264          &  & 0.671          & 0.634          & 0.798          & 0.855          \\
			& UCSV  & 0.227          & \textbf{0.223} & \textbf{0.229} & \textbf{0.226} &  & 0.765          & 0.731          & 0.737          & 0.723          \\
			& TVPSV & 0.581          & 0.593          & 0.596          & 0.582          &  & 0.950          & 0.972          & 0.978          & 0.968          \\
			& TVPQR & 0.230          & 0.262          & 0.240          & 0.260          &  & 0.690          & 0.661          & 0.673          & 0.745          \\
			& TVPDR & 0.223          & 0.247          & 0.237          & 0.243          &  & \textbf{0.647} & \textbf{0.601} & \textbf{0.651} & \textbf{0.541} \\ \hline
			\textit{QS: 95\%} & LR    & 0.167          & 0.159          & 0.173          & 0.174          &  & 0.439          & 0.587          & 0.668          & 0.614          \\
			& QR    & 0.208          & 0.180          & 0.174          & 0.232          &  & 0.558          & 0.533          & 0.699          & 1.034          \\
			& DR    & 0.181          & 0.183          & 0.174          & 0.181          &  & 0.507          & 0.487          & 0.506          & 0.917          \\
			& UCSV  & 0.185          & 0.189          & 0.186          & 0.187          &  & \textbf{0.297} & \textbf{0.298} & \textbf{0.306} & \textbf{0.303} \\
			& TVPSV & 0.582          & 0.591          & 0.598          & 0.591          &  & 0.979          & 0.930          & 0.987          & 1.001          \\
			& TVPQR & \textbf{0.161} & 0.176          & 0.176          & 0.169          &  & 0.426          & 0.478          & 0.456          & 0.463          \\
			& TVPDR & 0.166          & \textbf{0.149} & \textbf{0.144} & \textbf{0.166} &  & 0.448          & 0.449          & 0.435          & 0.617          \\ \hline
		\end{tabular}
		\vspace{0.1cm}
		
		\begin{minipage}{.9\linewidth} 
			\linespread{1}\footnotesize
			\textit{Notes}: This table reports the comparison results for headline inflation across various metrics and two distinct sample periods. Results are presented for forecasting horizons ($h=1,2,3,4$) and compared among different models. The best-performing values in each category are highlighted in bold.
		\end{minipage}
		\label{Tab: Headline}
	\end{table}

    Table \ref{Tab: Headline} summarizes the forecasting performance for headline inflation across all evaluation metrics. The TVPDR model ranks highest for both point and distribution accuracy in non-recession periods, delivering the lowest RMSE and CRPS across most short-to-medium horizons. The UCSV model is a strong competitor, providing the lowest RMSE at longer horizons and the lowest CRPS during recessions. The QS results further show that TVPDR leads in recessionary downside risk, whereas UCSV dominates recessionary upside risk. During recessions, the DR model exhibits a brief advantage in short-horizon point forecasts, but its performance is quickly surpassed by UCSV at longer horizons. The LR model delivers competitive distribution forecasts in stable periods and performs well for non-recession downside risk. TVPQR remains a strong performer for tail risks, particularly for non-recession upside risk. In contrast, QR and TVPSV produce consistently larger errors across most metrics and regimes, with TVPSV performing worst overall.

	\begin{table}[H]
		\footnotesize
		\centering
		\caption{Forecasting Performance Comparison for Core CPI Inflation}
	    \begin{tabular}{lllllllllll}
	    	\hline
	    	\multicolumn{2}{l}{}      & \multicolumn{4}{c}{Non-Recession Periods}                         &                         & \multicolumn{4}{c}{Recession Periods}                             \\ \cline{3-6} \cline{8-11} 
	    	\multicolumn{2}{l}{}      & $h=1$          & $h=2$          & $h=3$          & $h=4$          &                         & $h=1$          & $h=2$          & $h=3$          & $h=4$          \\ \hline
	    	\textit{RMSE}     & LR    & 0.562          & 0.606          & 0.665          & 0.644          &                         & 1.779          & 1.798          & 1.813          & 2.098          \\
	    	& QR    & 0.641          & 0.683          & 0.604          & 0.653          &                         & 2.629          & 2.794          & 2.651          & 3.182          \\
	    	& DR    & 0.566          & 0.616          & 0.636          & 0.610          &                         & 1.830          & 1.802          & 1.748          & 2.023 \\
	    	& UCSV  & \textbf{0.509} & \textbf{0.501} & \textbf{0.514} & \textbf{0.053} & {\color[HTML]{4472C4} } & 1.869          & 1.846          & 1.918          & \textbf{1.877} \\
	    	& TVPSV & 0.609          & 0.638          & 0.700          & 0.639          &                         & 2.310          & 2.014          & 1.830          & 3.182          \\
	    	& TVPQR & 0.547          & 0.565          & 0.630          & 0.643          &                         & 2.408          & 2.132          & 2.216          & 3.264          \\
	    	& TVPDR & 0.591          & 0.579          & 0.590          & 0.597          &                         & \textbf{1.470} & \textbf{1.470} & \textbf{1.379} & 2.132          \\ \hline
	    	\textit{CRPS}     & LR    & 0.346          & 0.356          & 0.380          & 0.369          &                         & 0.952          & 1.002          & 0.992          & 1.149          \\
	    	& QR    & 0.371          & 0.391          & 0.361          & 0.371          &                         & 1.084          & 1.122          & 1.150          & 1.391          \\
	    	& DR    & 0.340          & 0.348          & 0.347          & 0.356          &                         & 1.253          & 1.485          & 0.939          & 1.367          \\
	    	& UCSV  & \textbf{0.302} & \textbf{0.299} & \textbf{0.302} & \textbf{0.300} & {\color[HTML]{4472C4} } & 0.977          & 0.958          & 0.993          & \textbf{0.971} \\
	    	& TVPSV & 1.575          & 1.614          & 1.611          & 1.580          &                         & 1.959          & 1.966          & 2.000          & 2.041          \\
	    	& TVPQR & 0.331          & 0.325          & 0.364          & 0.366          &                         & 1.058          & 0.982          & 1.004          & 1.122          \\
	    	& TVPDR & 0.334          & 0.332          & 0.344          & 0.340          &                         & \textbf{0.792} & \textbf{0.887} & \textbf{0.779} & 1.226          \\ \hline
	    	\textit{QS: 5\%}  & LR    & \textbf{0.060} & \textbf{0.063} & 0.095          & 0.098          &                         & 0.235          & 0.220          & 0.237          & 0.228          \\
	    	& QR    & 0.084          & 0.077          & 0.085          & 0.076          &                         & 0.299          & 0.253          & 0.248          & 0.276          \\
	    	& DR    & 0.095          & 0.085          & 0.070          & 0.077          &                         & 0.293          & 0.302          & 0.274          & 0.252          \\
	    	& UCSV  & 0.074          & 0.074          & 0.074          & 0.075          & {\color[HTML]{4472C4} } & 0.314          & 0.302          & 0.317          & 0.306          \\
	    	& TVPSV & 0.553          & 0.556          & 0.562          & 0.551          &                         & 0.676          & 0.653          & 0.677          & 0.688          \\
	    	& TVPQR & 0.073          & 0.080          & 0.075          & \textbf{0.072} &                         & 0.245          & 0.248          & 0.288          & 0.252          \\
	    	& TVPDR & 0.065          & 0.066          & \textbf{0.065} & 0.074          &                         & \textbf{0.231} & \textbf{0.209} & \textbf{0.228} & \textbf{0.214} \\ \hline
	    	\textit{QS: 95\%} & LR    & 0.068          & 0.078          & 0.078          & 0.079          &                         & 0.253          & 0.256          & 0.293          & 0.428          \\
	    	& QR    & 0.074          & 0.082          & 0.074          & 0.077          &                         & 0.231          & 0.250          & 0.263          & 0.323          \\
	    	& DR    & 0.083          & 0.081          & 0.085          & 0.083          &                         & 0.199          & 0.179          & 0.172          & 0.316          \\
	    	& UCSV  & 0.058          & \textbf{0.058} & \textbf{0.057} & \textbf{0.058} & {\color[HTML]{4472C4} } & 0.201          & 0.209          & 0.209          & 0.212          \\
	    	& TVPSV & 0.541          & 0.555          & 0.556          & 0.546          &                         & 0.651          & 0.660          & 0.659          & 0.662          \\
	    	& TVPQR & 0.058          & 0.066          & 0.073          & 0.080          &                         & \textbf{0.092} & 0.177          & 0.143          & \textbf{0.167} \\
	    	& TVPDR & \textbf{0.055} & 0.066          & 0.078          & 0.075          &                         & 0.155          & \textbf{0.171} & \textbf{0.136} & 0.223          \\ \hline
	    \end{tabular}
	    \vspace{0.1cm}
	    
		\begin{minipage}{.9\linewidth} 
			\linespread{1}\footnotesize
			\textit{Notes}: This table reports the comparison results for core inflation across various metrics and two distinct sample periods. Results are presented for forecasting horizons ($h=1,2,3,4$) and compared among different models. The best-performing values in each category are highlighted in bold.
		\end{minipage}
		\label{Tab: Core}
	\end{table}

	Table \ref{Tab: Core} reports the corresponding results for core inflation and also reveals a clear separation in model dominance across economic regimes. In non-recession periods, the UCSV model is the best overall performer, exhibiting superior point, distribution, and tail forecast accuracy. It achieves the lowest RMSE and CRPS across nearly all non-recession horizons and delivers the most accurate forecasts for upside risk at longer horizons. In contrast, the TVPDR model dominates during recessions, providing the lowest RMSE, CRPS, and downside tail forecasts for horizons $h=1, 2, 3$. The UCSV model remains competitive in downturns, taking the lead at $h=4$ for both RMSE and CRPS, reflecting strong long-horizon robustness. TVPQR also exhibits solid performance, frequently ranking second in RMSE and delivering accurate forecasts for recessionary upside risk. The simpler LR and DR models offer moderate performance, with LR performing notably well for non-recession downside risk. The QR model consistently produces larger errors. Most notably, the TVPSV model performs the worst across all core inflation metrics, underscoring its limited ability to model the dynamics of core inflation.

  	Finally, we evaluate the probabilistic calibration of the predictive densities using the probability integral transform (PIT), defined as the predictive cumulative distribution function evaluated at the true realization. In a perfectly calibrated model, the PITs are independently and identically distributed uniform variates. To assess calibration across all benchmark models, we adopt the statistical framework proposed by \cite{rossi2014evaluating}, which tests for both uniformity and independence in the PIT series. We employ three tests for uniformity: the Kolmogorov-Smirnov (KS), the Anderson-Darling (AD), and the Doornik-Hansen (DH) tests.\footnote{The KS test is sensitive to overall deviations from uniformity, the AD test places greater weight on tail deviations (near 0 or 1), and the DH test is sensitive to skewness and kurtosis in the transformed series.} Independence is assessed using the Ljung-Box (LB) test applied to the first (LB1, mean) and second (LB2, variance) central moments of the PIT series. Table \ref{Tab: PITs} presents the $p$-values of the above statistical tests applied to the PITs for both headline and core CPI inflation forecasts at horizons $h = 1$ and $h = 4$.  As is standard practice, $p$-values exceeding the 5\% significance level suggest no evidence against the null hypothesis of proper calibration, and are highlighted in bold in the table.

	\begin{table}[H]
		\centering
		\footnotesize
		\caption{Probability Integral Transforms (PITs) Test Results}
		\begin{tabular}{llllllllllll}
			\hline
			& \multicolumn{5}{l}{$h=1$}                                                          &           & \multicolumn{5}{l}{$h=4$}                                                          \\ \cline{2-6} \cline{8-12} 
			& KS             & AD             & DH             & LB1            & LB2            &           & KS             & AD             & DH             & LB1            & LB2            \\ \hline
			\multicolumn{12}{l}{\textit{Headline CPI Inflation}}                                                                                                                                        \\
			LR    & \textbf{0.956} & \textbf{0.462} & \textbf{0.822} & \textbf{0.213} & \textbf{0.078} &           & \textbf{0.604} & \textbf{0.058} & \textbf{0.682} & 0.000          & 0.002          \\
			QR    & \textbf{0.104} & 0.001          & \textbf{0.988} & \textbf{0.487} & \textbf{0.668} &           & \textbf{0.072} & 0.000          & \textbf{0.970} & 0.000          & \textbf{0.056} \\
			DR    & 0.032          & 0.000          & \textbf{0.886} & \textbf{0.553} & \textbf{0.475} &           & 0.023          & 0.000          & \textbf{0.889} & 0.000          & 0.032          \\
			UCSV  & \textbf{0.577} & \textbf{0.492} & \textbf{0.876} & \textbf{0.765} & \textbf{0.171} & \textbf{} & \textbf{0.745} & \textbf{0.552} & \textbf{0.899} & \textbf{0.825} & \textbf{0.382} \\
			TVPSV & 0.000          & 0.000          & \textbf{0.161} & \textbf{0.493} & \textbf{0.771} &           & 0.000          & 0.000          & \textbf{0.340} & \textbf{0.402} & \textbf{0.423} \\
			TVPQR & \textbf{0.364} & \textbf{0.116} & \textbf{0.358} & \textbf{0.107} & \textbf{0.200} &           & \textbf{0.371} & \textbf{0.174} & \textbf{0.193} & 0.044          & \textbf{0.142} \\
			TVPDR & \textbf{0.922} & \textbf{0.750} & \textbf{0.819} & \textbf{0.467} & \textbf{0.982} & \textbf{} & \textbf{0.876} & \textbf{0.695} & \textbf{0.938} & \textbf{0.432} & \textbf{0.755} \\ \hline
			\multicolumn{12}{l}{\textit{Core CPI Inflation}}                                                                                                                                            \\
			LR    & \textbf{0.289} & \textbf{0.157} & \textbf{0.350} & \textbf{0.370} & \textbf{0.103} &           & 0.001          & 0.000          & \textbf{0.162} & 0.000          & 0.000          \\
			QR    & 0.020          & 0.000          & \textbf{0.983} & \textbf{0.269} & \textbf{0.225} &           & 0.002          & 0.000          & \textbf{0.879} & 0.001          & 0.001          \\
			DR    & 0.004          & 0.000          & \textbf{0.893} & \textbf{0.232} & \textbf{0.938} &           & 0.000          & 0.000          & \textbf{0.871} & 0.028          & \textbf{0.115} \\
			UCSV  & \textbf{0.885} & \textbf{0.948} & \textbf{0.915} & \textbf{0.531} & 0.021          & \textbf{} & \textbf{0.967} & \textbf{0.954} & \textbf{0.911} & \textbf{0.539} & 0.032          \\
			TVPSV & 0.000          & 0.000          & 0.000          & 0.013          & 0.004          &           & 0.000          & 0.000          & 0.000          & \textbf{0.503} & 0.000          \\
			TVPQR & \textbf{0.296} & \textbf{0.480} & \textbf{0.751} & \textbf{0.304} & \textbf{0.950} & \textbf{} & \textbf{0.651} & \textbf{0.562} & \textbf{0.556} & \textbf{0.089} & \textbf{0.060} \\
			TVPDR & \textbf{0.542} & \textbf{0.267} & \textbf{0.398} & \textbf{0.207} & \textbf{0.401} &           & \textbf{0.371} & \textbf{0.117} & \textbf{0.351} & \textbf{0.067} & 0.000          \\ \hline
		\end{tabular}
		\vspace{0.1cm}				
		\begin{minipage}{.95\linewidth} 
			\linespread{1}\footnotesize
			\textit{Notes}: This table reports p-values from diagnostic tests applied to the PITs of the predictive densities. The tests include: Kolmogorov-Smirnov (KS), Anderson-Darling (AD), and Doornik-Hansen (DH) for uniformity; and Ljung-Box tests for the mean (LB1) and variance (LB2) for independence. Following \cite{rossi2014evaluating}, p-values exceeding 5\% (indicating no evidence against the null hypothesis) are highlighted in bold, suggesting the model satisfies the corresponding statistical properties.
		\end{minipage}
		\label{Tab: PITs}
	\end{table}

	For the short-run horizon $h=1$, the flexible models (UCSV, TVPQR, TVPDR), alongside the static LR, demonstrated good calibration across both headline and core inflation, satisfying all uniformity and independence tests. In sharp contrast, static models (QR, DR) and the TVPSV consistently failed uniformity-based tests, indicating misspecified predictive densities, with TVPSV performing particularly poorly on core inflation. At the longer $h=4$ horizon, the advantages of the TVPDR and UCSV models became pronounced, demonstrating the strongest overall performance and maintaining high $p$-values across all diagnostic checks for headline inflation. For core inflation at $h=4$, only the TVPDR, UCSV, and TVPQR models maintained acceptable calibration, while all other benchmarks showed significant deficiencies in uniformity and independence.

   Overall, the empirical evidence suggests that models combining time-varying parameters with full distributional flexibility are optimally suited not only for capturing evolving inflation dynamics but also for delivering reliable forecasts across the entire probability space.

	\subsection{Inflation Risks Analysis}
		
  Effective monetary policy is inherently risk-management-oriented and requires assessing inflation outcomes relative to an explicit numerical target or target range. Deviations on either side of the target can entail substantial macroeconomic costs, with the consequences often differing across downside and upside risks. Traditional quantile-based risk measures, while informative, may not align well with central banks' asymmetric loss functions and can miss economically relevant changes in risk when the predictive distribution shifts in ways not reflected by fixed quantiles. Our TVPDR model provides a more flexible and policy-relevant framework by evaluating inflation risks across the full predictive distribution relative to a specified target range.

\medskip

Focusing on one-quarter-ahead forecasting ($h=1$), we assess the model's ability to forecast the risks that inflation falls substantially above or below the target range and to identify the key drivers of these risks, using the risk measures of \cite{kilian2007quantifying} introduced in subsection~\ref{subsec: risk measure}.

	\subsubsection{Out-of-sample Inflation Risks Forecasting}
	
	We initiate the analysis of inflation risks using simplified risk measures where the target loss parameters are set to zero ($\alpha=\gamma=0$). Under this specification, the risk measures simplify to the probability of inflation falling below or exceeding a specified target range.

\begin{figure}[H]
		\captionsetup[subfigure]{aboveskip=-2pt,belowskip=0pt}
		\centering
		\caption{Probabilities of Deflation and Excessive Inflation Risks ($\alpha=\gamma=0$)}
		\begin{subfigure}[b]{0.48\textwidth}
			\centering
			\caption{ Headline Inflation }	
			\label{fig: Risk0-Head}
			\includegraphics[width=1\textwidth, height=0.75\textwidth]{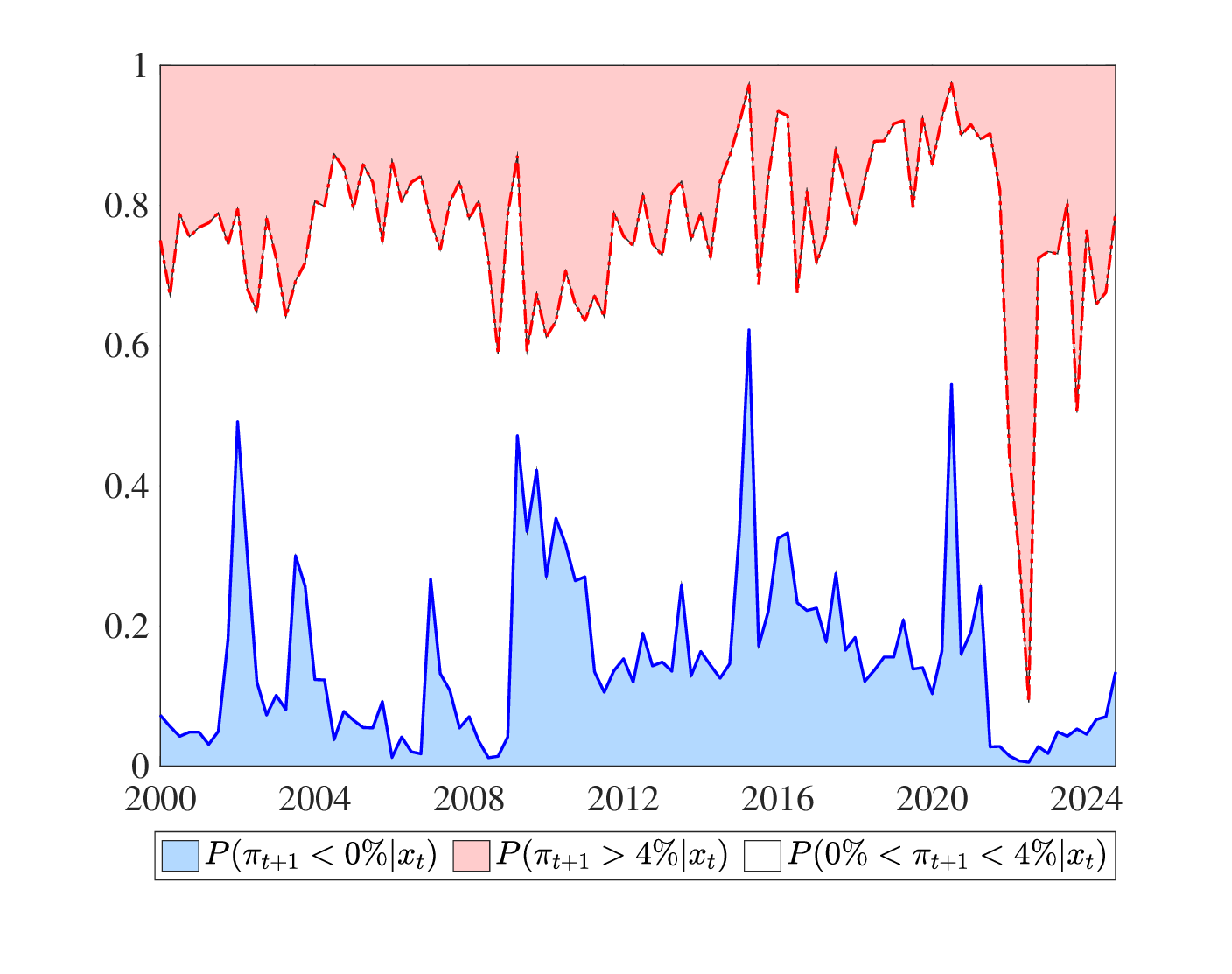}
		\end{subfigure}
		\begin{subfigure}[b]{0.48\textwidth}
			\centering
			\caption{Core Inflation }	
			\label{fig: Risk0-Core}
			\includegraphics[width=1\textwidth, height=0.75\textwidth]{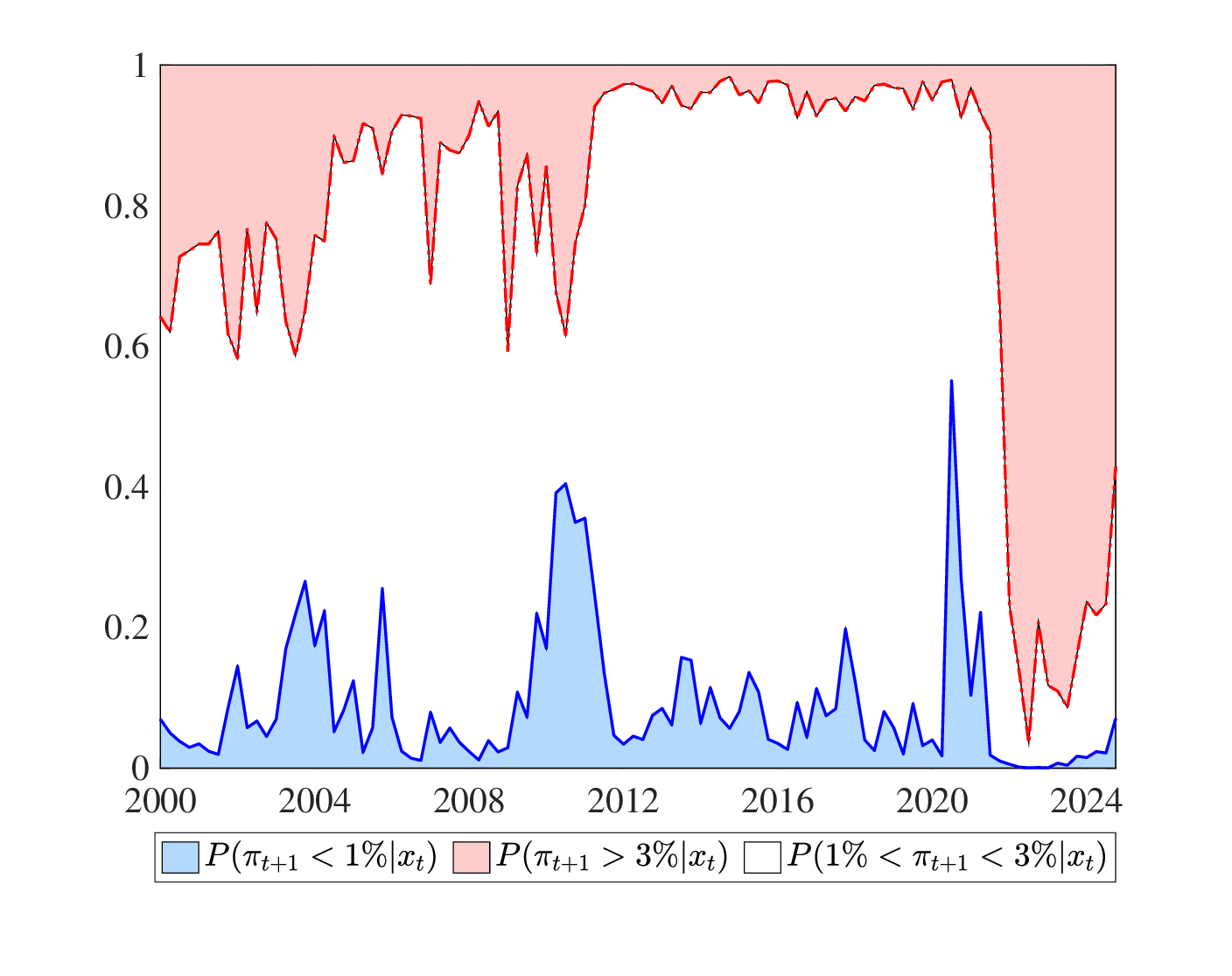}
		\end{subfigure}
		\begin{minipage}{.9\linewidth} 
			\linespread{1}\footnotesize
			\textit{Notes}: This three-region figure shows the probabilities of inflation falling within or outside the target range, as measured by the risk measures when $\alpha=\gamma=0$. Panel (a) presents results for headline inflation with a target range of $[0\%, 4\%]$, and Panel (b) for core inflation with a range of $[1\%, 3\%]$. In each panel, the solid blue line represents the probability of deflation, and the dashed red line indicates the probability of excessive inflation. 
		\end{minipage}
		\label{fig: Risks}
	\end{figure}

    Figure $\ref{fig: Risks}$(a) reports the time-series evolution of these probabilities for headline inflation, relative to a baseline target range of $[0\%, 4\%]$. The TVPDR model effectively captures both deflation risk ($\pi_{t+1} < 0\%$) and excessive inflation risk ($\pi_{t+1} > 4\%$). We observe that the deflation risk probability spikes during major economic downturns, surpassing a probability of $0.5$ during certain periods, but remains contained below $0.2$ during stable economic periods. Prior to the Great Recession, the probability of excessive inflation risk hovered near $0.3$ before declining below $0.1$ by 2021. However, the post-pandemic supply disruptions and robust demand triggered a sharp surge in 2022, with the probability peaking near $0.9$. This risk has since moderated to approximately $0.3$.

	 Figure \ref{fig: Risks}(b) presents results for core inflation, focusing on tail risks relative to a narrower target range of $[1\%, 3\%]$, reflecting its lower volatility. The model identifies elevated deflation risk probability during downturns, reaching a level of $0.4$ during the Great Recession and $0.6$ during the pandemic shock, while remaining below $0.1$ in stable periods. The excessive inflation risk probability stays around $0.1$ prior to 2021, rises sharply to a probability of $0.9$ in 2022, and gradually declines to $0.6$ by the end of 2024. The excessively high risk for core inflation has remained persistently elevated post-pandemic, unlike the more rapid moderation seen in headline inflation. This suggests that core measures more effectively capture underlying price pressures, providing a crucial distinction for policy assessment.

\begin{figure}[H]
		\captionsetup[subfigure]{aboveskip=-2pt,belowskip=0pt}
		\centering
		\caption{Expected Deflation and Excessive Inflation Risks ($\alpha=\gamma=1$)}
		\begin{subfigure}[b]{0.48\textwidth}
			\centering
			\caption{Headline Inflation }	
			\label{fig: Risk1-Head}
			\includegraphics[width=1\textwidth, height=0.7\textwidth]{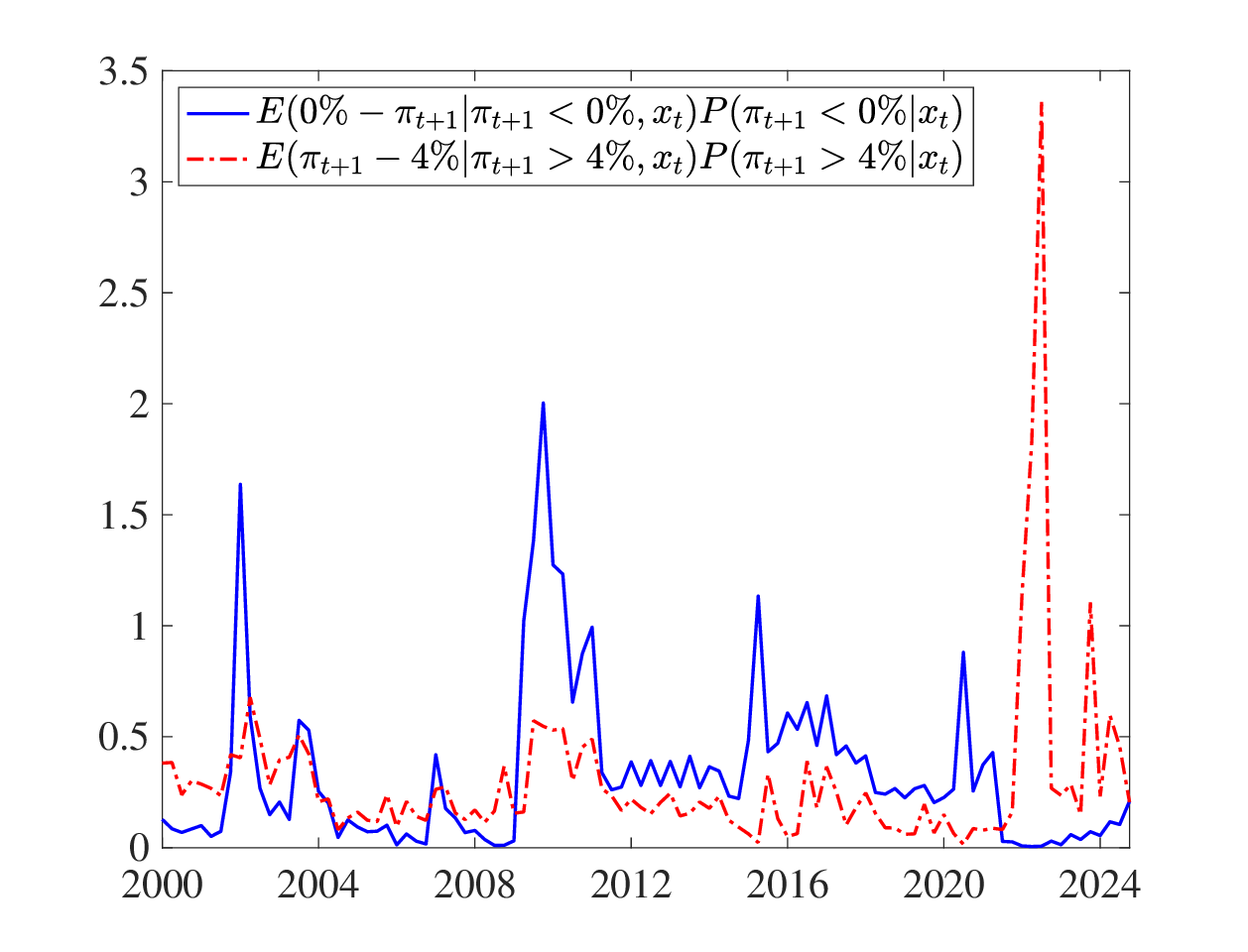}
		\end{subfigure}
		\begin{subfigure}[b]{0.48\textwidth}
			\centering
			\caption{Core Inflation }	
			\label{fig: Risk1-Core}
			\includegraphics[width=1\textwidth, height=0.7\textwidth]{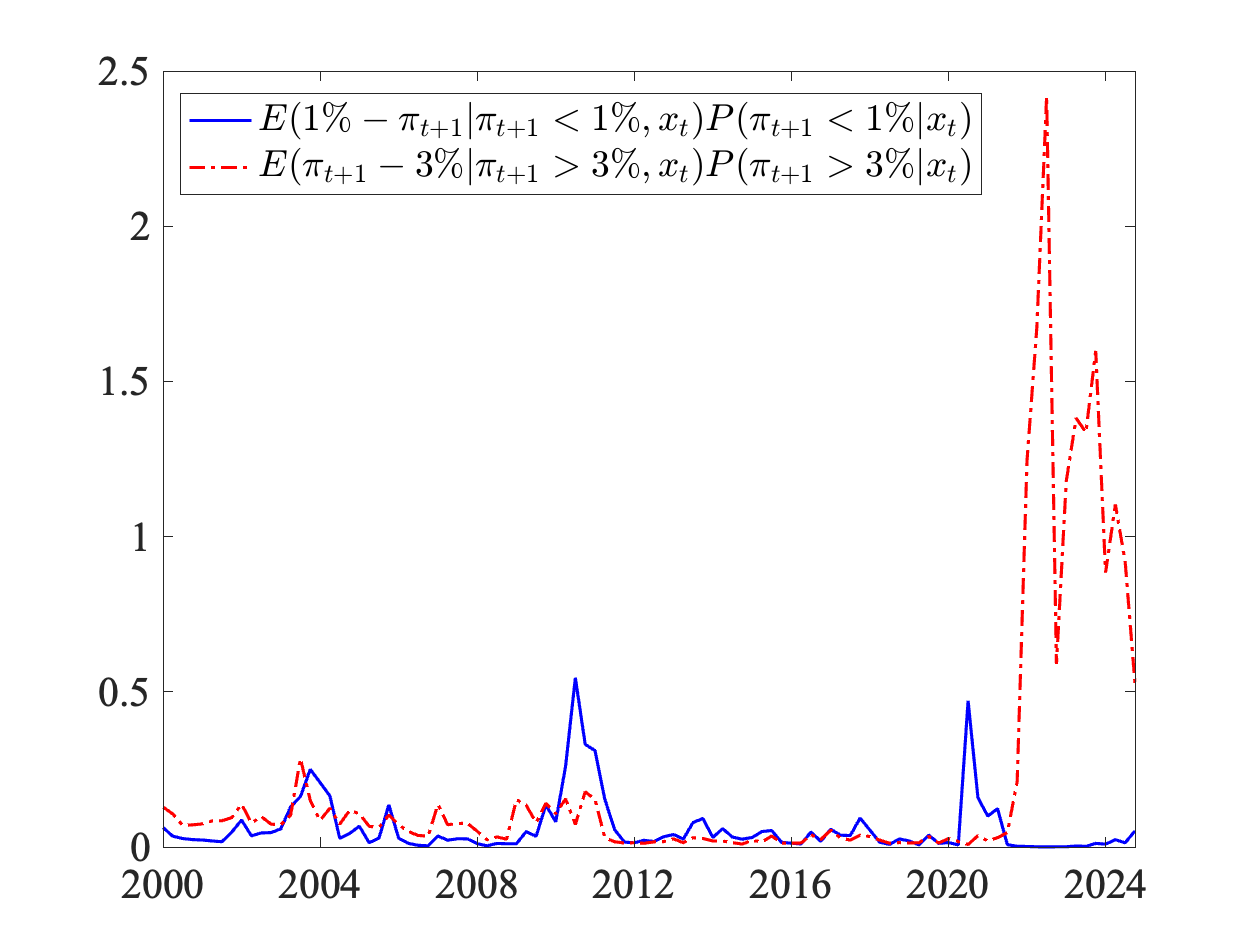}
		\end{subfigure}
		\begin{minipage}{.9\linewidth} 
			\linespread{1}\footnotesize
			\textit{Notes}: This figure shows the expected deflation and excessive inflation relative to a specified target range, as measured by the risk measures when $\alpha=\gamma=1$. Refer to Figure \ref{fig: Risks} for more details.
		\end{minipage}
			\label{fig: Risks1}
	\end{figure}

    However, probabilities alone may not fully capture the policy-relevant nature of inflation risks. Economic agents and policymakers typically focus not only on the likelihood of inflation breaching the target zone but also on the expected magnitude of such deviations. In this context, risk measures with $\alpha = \gamma = 1$ offer a more informative assessment by jointly reflecting both the probability and the severity of tail events. The corresponding time series are presented in Figure $\ref{fig: Risks1}$.

	  For headline inflation shown in Figure \ref{fig: Risks1}(a), the expected excessive inflation rises sharply in the post-pandemic period. This trajectory reflects a confluence of high probability of breaching the upper bound and substantial potential overshooting, consistent with the observed multi-decade peak in realized inflation. In contrast, the expected deflation risk is tightly concentrated in three distinct episodes: the early 2000s recession, the 2008-2009 Great Recession, and the 2020 pandemic shock. Crucially, in periods outside these major crises, both risks remain negligible once the magnitude of potential deviations is incorporated. This suggests that the $\alpha=\gamma=0$ probability-based measures may significantly overstate inflation risks during normal economic conditions. From Figure \ref{fig: Risks1}(b) for core inflation, the results show minor deflation risks only during the Great Recession and pandemic periods, showing that these events appear less severe when magnitude is incorporated. It is particularly noteworthy that the risks from expected excessive inflation indicate a heightened concern about the potential for higher inflation following the pandemic. Outside of crisis periods, both deflation and excessive inflation risks become insignificant when accounting for deviation magnitudes. 
	
     Overall, the combination of probability and magnitude measures provides more policy-relevant signals than focusing on probability alone. Although the TVPDR model exhibits slight delays in detecting certain extreme inflation outcomes relative to the full realized data, it effectively captures major shifts in tail risks, particularly during periods of acute economic distress, offering timely signals for preemptive monetary policy action.
	
	\subsubsection{Drivers of Inflation Risks via Shapley Value Decomposition}
	
	To identify the drivers of time-varying inflation risks, we employ Shapley values to decompose the model-implied risk measures into contributions from individual predictors \citep{shapley1953value, strumbelj2010efficient}. This approach has recently been applied to inflation risk analysis by \cite{lenza2025density}, within a QR random forest framework.
		
	In the semiparametric TVPDR model, the effects of macroeconomic variables on inflation risks are inherently nonlinear and state-dependent. Shapley values provide a locally accurate and additive decomposition of the predicted inflation risk into marginal contributions attributable to each predictor. This decomposition quantifies how each variable shifts inflation risks relative to a historical-average baseline, offering a transparent and economically interpretable assessment of the sources of deflation and excessive inflation risks.

\begin{figure}[H]
    	\captionsetup[subfigure]{aboveskip=-2pt,belowskip=0pt}
    	\centering
    	\caption{Contribution of risk factors to Inflation Risks}
    	\begin{subfigure}[b]{0.45\textwidth}
    		\centering
    		\caption{Deflation: $P(\pi_{t+1}<0\%\mid x_t)$}	
    		\label{fig: Shapley-a}
    		\includegraphics[width=1\textwidth, height=0.7\textwidth]{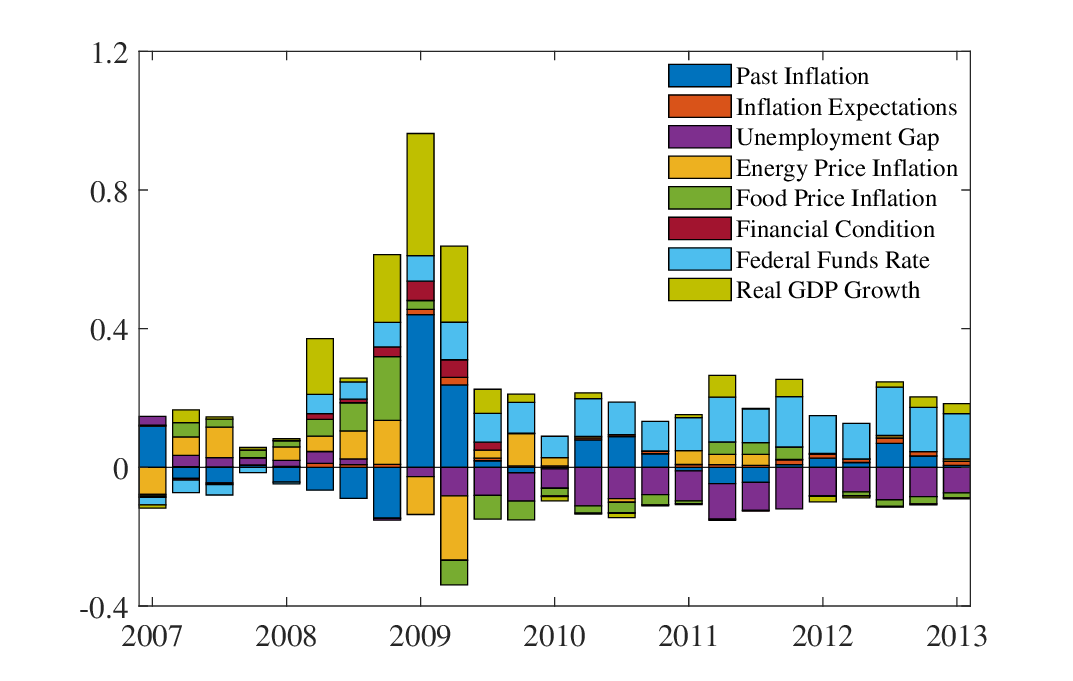}    		
    	\end{subfigure}
    	\begin{subfigure}[b]{0.45\textwidth}
    		\centering
    		\caption{Excessive Inflation: $P(\pi_{t+1}>4\%\mid x_t)$}	
    		\includegraphics[width=1\textwidth, height=0.7\textwidth]{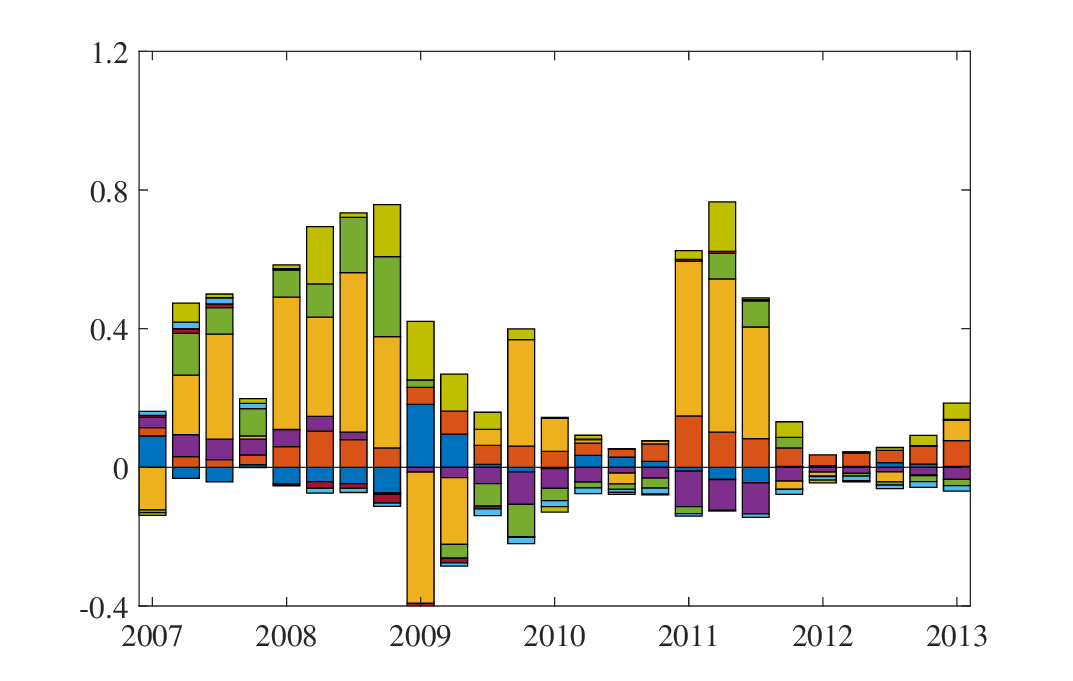}
    	\end{subfigure}
    	\vspace{0.2cm}
    	\begin{subfigure}[b]{0.45\textwidth}
    		\centering
    		\caption{Deflation: $P(\pi_{t+1}<0\%\mid x_t)$}	
    		\includegraphics[width=1\textwidth, height=0.7\textwidth]{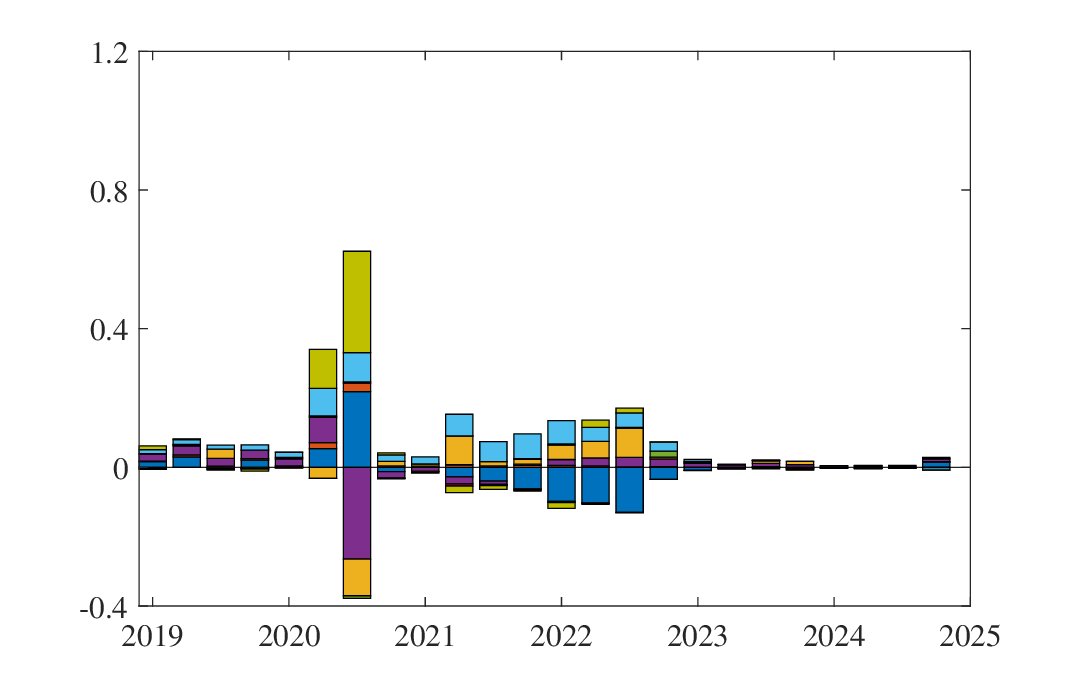}
    	\end{subfigure}
    	\begin{subfigure}[b]{0.45\textwidth}
    		\centering
    		\caption{Excessive Inflation: $P(\pi_{t+1}>4\%\mid x_t)$}	
    		\includegraphics[width=1\textwidth, height=0.7\textwidth]{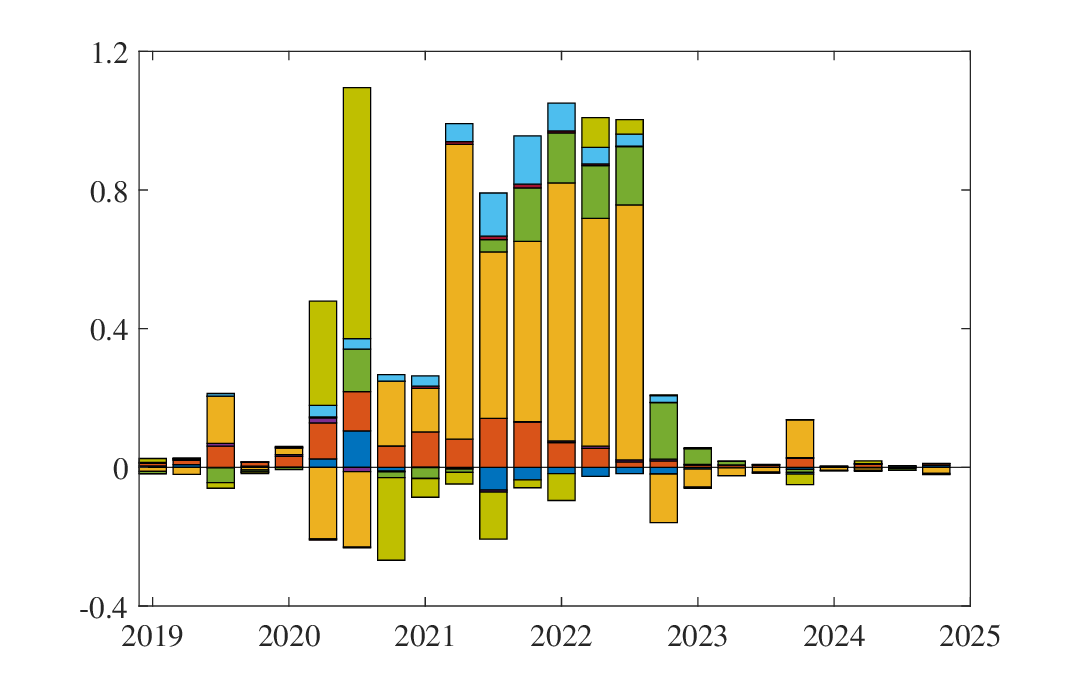}
    	\end{subfigure}
    	\begin{minipage}{.9\linewidth} 
    		\linespread{1}\footnotesize
    		\textit{Notes}: This figure displays stacked bar plots of the Shapley value decomposition, illustrating each economic variable’s contribution to the predicted probabilities of deflation and excessive inflation risk. The analysis focuses on two extended periods associated with heightened inflation-related risks: the Great Recession and its aftermath (2007–2013), the pandemic and subsequent recovery (2019–2025).
    	\end{minipage}
    	\label{fig: Shapley}
    \end{figure}

	Figure \ref{fig: Shapley} summarizes the decomposition results for deflation and excessive inflation risks across two key periods: the Great Recession and its aftermath (2007–2013), and the pandemic and recovery (2019–2025). The stacked bars in each panel illustrate the sign and magnitude of the contribution of each variable relative to the baseline probability. From Figure \ref{fig: Shapley}(a), the increase in deflation risk during the 2008 financial crisis is mainly driven by sharp drops in real GDP growth and falling energy and food price inflation. From 2009 onward, the model attributes much of the disinflationary pressure to lagged inflation, which reflects the lingering effects of earlier price shocks, especially from energy. During the zero lower bound (ZLB) period, the low federal funds rate signals limited policy space and weak inflation momentum, adding to deflation risk. However, as unemployment gradually fell after 2009, an improving labor market helped offset this, leading to relatively low deflation probabilities during much of the ZLB period. According to Figure \ref{fig: Shapley}(b), the analysis indicates that the modest, temporary spikes in excessive inflation risk around 2008 and 2011 were primarily driven by acute, exogenous energy price inflation surges, consistent with global oil price spikes. The sharp decline in energy prices at the end of 2008 then played a dominant role in reducing the predicted excessive inflation risk, demonstrating the transitory nature of these supply shocks.

    The pandemic period presents a different pattern. In the early stages of the pandemic in 2020, Figure \ref{fig: Shapley}(c) shows that the sharp decline in real GDP growth raised deflation risk through demand-side channels. At the same time, Figure \ref{fig: Shapley}(d) illustrates that the unique nature of the pandemic shock, marked by simultaneous supply disruptions and aggressive policy responses, also elevated excessive inflation risk, driven by heightened uncertainty and constrained production capacity. Furthermore, from Figure \ref{fig: Shapley}(c), real GDP growth and lagged inflation continued to exert upward pressure on deflation risk early in the pandemic. However, the surge in unemployment did not lower inflation as traditional models predict, suggesting a temporary breakdown or muting of the usual PC relationship. Beginning in 2021, Figure \ref{fig: Shapley}(d) shows that energy price inflation became the main driver of excessive inflation risk, alongside smaller contributions from food prices, inflation expectations, and low interest rates. This marks a shift from demand-driven deflation to supply-side inflation pressures during the recovery.
    
    Overall, the Shapley value decomposition provides clear evidence that deflation and excessive inflation risks arise from distinct mechanisms. Deflation risk is primarily associated with demand-side weakness and inflation inertia, while excessive inflation risk stems mainly from supply-side shocks, particularly energy and food prices. Real GDP growth acts as a dual-role indicator, driving downside risk during contractions and signaling potential overheating risk during expansions. This asymmetric distribution of risk drivers highlights the necessity of using a time-varying, interpretable distributional model like TVPDR for policy assessment.

	 \subsubsection{Sensitivity of Inflation Risks via Scenario Analysis} 
	 
	We explore the sensitivity of inflation tail risks to counterfactual changes in key macroeconomic variables over the recent period (2015:Q1 to 2024:Q4), focusing on energy price inflation (EPI) and the unemployment rate. Using the TVPDR framework, we simulate hypothetical deviations from observed paths to assess how changes in these variables affect the probabilities of deflation risk and excessive inflation risk.

     \begin{figure}[H]
    	\captionsetup[subfigure]{aboveskip=-2pt,belowskip=0pt}
    	\centering
    	\caption{Impact of Energy Price Inflation on Inflation Risk Probabilities}
    	\begin{subfigure}[b]{0.5\textwidth}
    		\centering
    		\caption{DR: $P(\pi_{t+1}<0\%\mid x_t)$}				 \includegraphics[width=1\textwidth, height=0.7\textwidth]{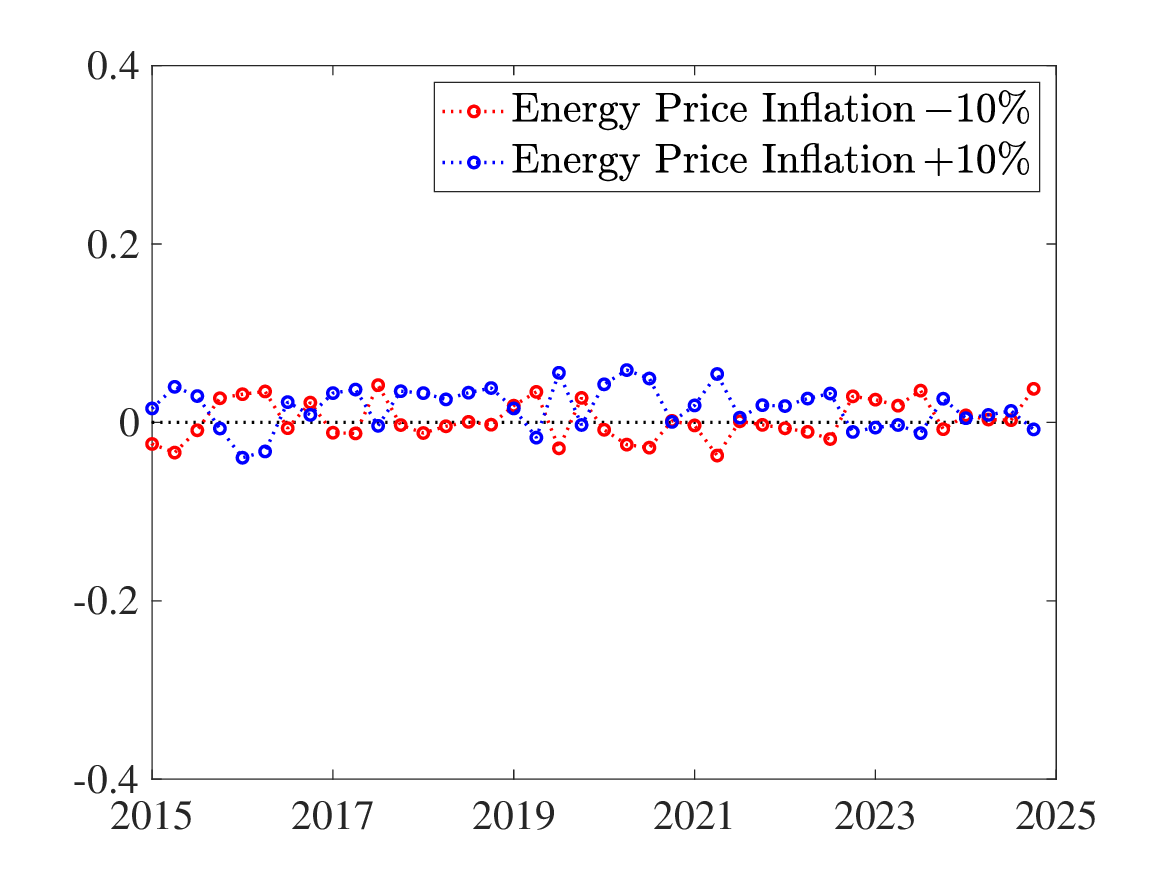}
    	\end{subfigure}
    	\hspace{-0.5cm}
    	\begin{subfigure}[b]{0.5\textwidth}
    		\centering
    		\caption{EIR: $P(\pi_{t+1}>4\%\mid x_t)$}	
    		\includegraphics[width=1\textwidth, height=0.7\textwidth]{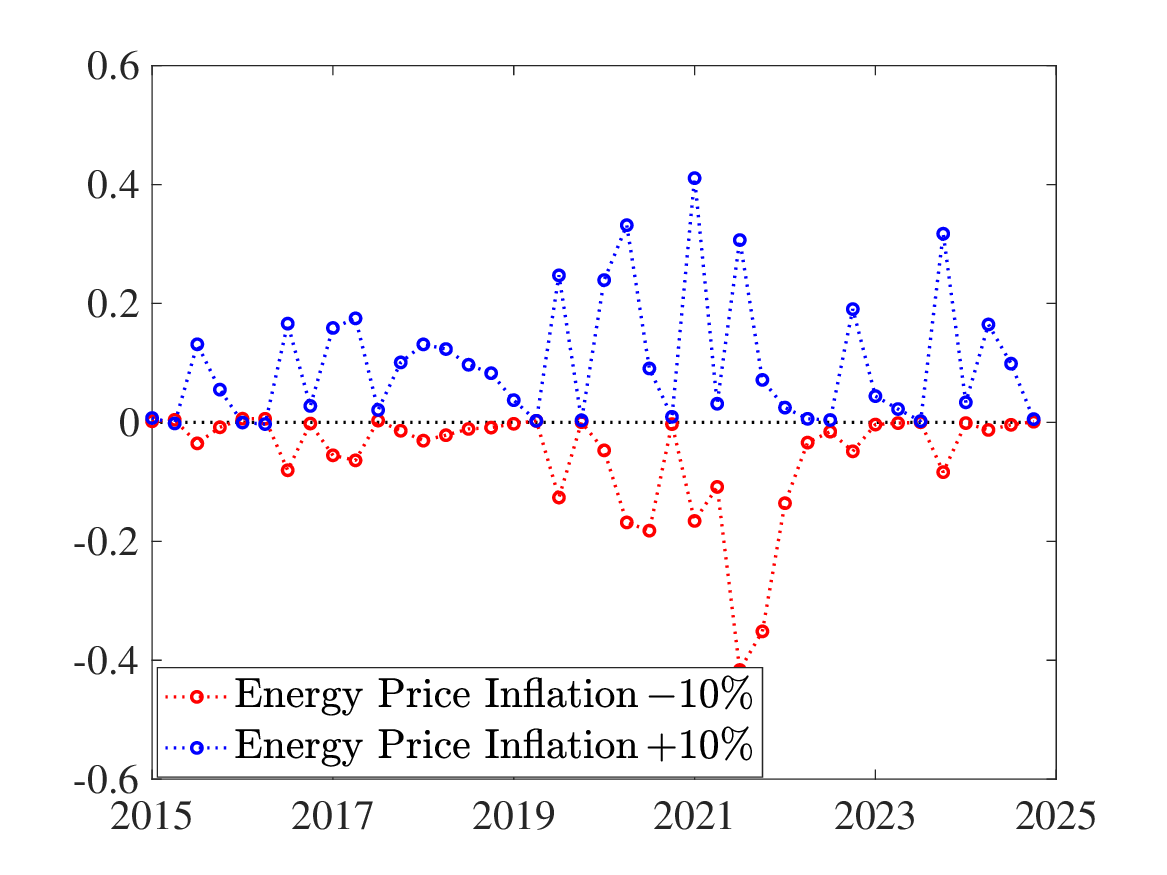}  	
    	\end{subfigure}
    	\begin{minipage}{.9\linewidth} 
    		\linespread{1}\footnotesize
    		\textit{Notes}: This figure shows the change in deflation and excessive inflation risk probabilities resulting from a 10\% decrease (red) and a 10\% increase (blue) in energy price inflation, applied each quarter. Markers indicate the change in probability each quarter, with positive values representing increases and negative values representing decreases.
    	\end{minipage}
    	\label{fig: Counter-EPI}
    \end{figure}

    Figure~\ref{fig: Counter-EPI} reports the effects of quarterly $\pm10$ percentage-point changes in energy price inflation on the probabilities of deflation and excessive inflation. The results indicate that variations in EPI exert a negligible influence on deflation risk throughout the sample. By contrast, the sensitivity to EPI on excessive inflation risk is notable and highly time-dependent. A $10$ percentage-point increase in EPI consistently raises the probability of excessive inflation, while a decrease lowers it. Crucially, the upper-tail risk response is asymmetric and time-varying. In most periods, increases in EPI generate a stronger rise in upper-tail risk than the corresponding decline reduces it. However, during the intense inflationary environment of 2021-2022, a reduction in EPI had a comparatively larger dampening effect on excessive inflation risk. These patterns underscore the role of energy price dynamics as a systematic and time-sensitive driver of upside inflation risk, particularly in periods of strong commodity price shocks and heightened inflation volatility.

      \begin{figure}[H]
     	\captionsetup[subfigure]{aboveskip=-2pt,belowskip=0pt}
     	\centering
     	\caption{Impact of Unemployment Rate on Inflation Risk Probabilities}
     	\begin{subfigure}[b]{0.5\textwidth}
     		\centering
     		\caption{DR: $P(\pi_{t+1}<0\%\mid x_t)$}				 \includegraphics[width=1\textwidth, height=0.7\textwidth]{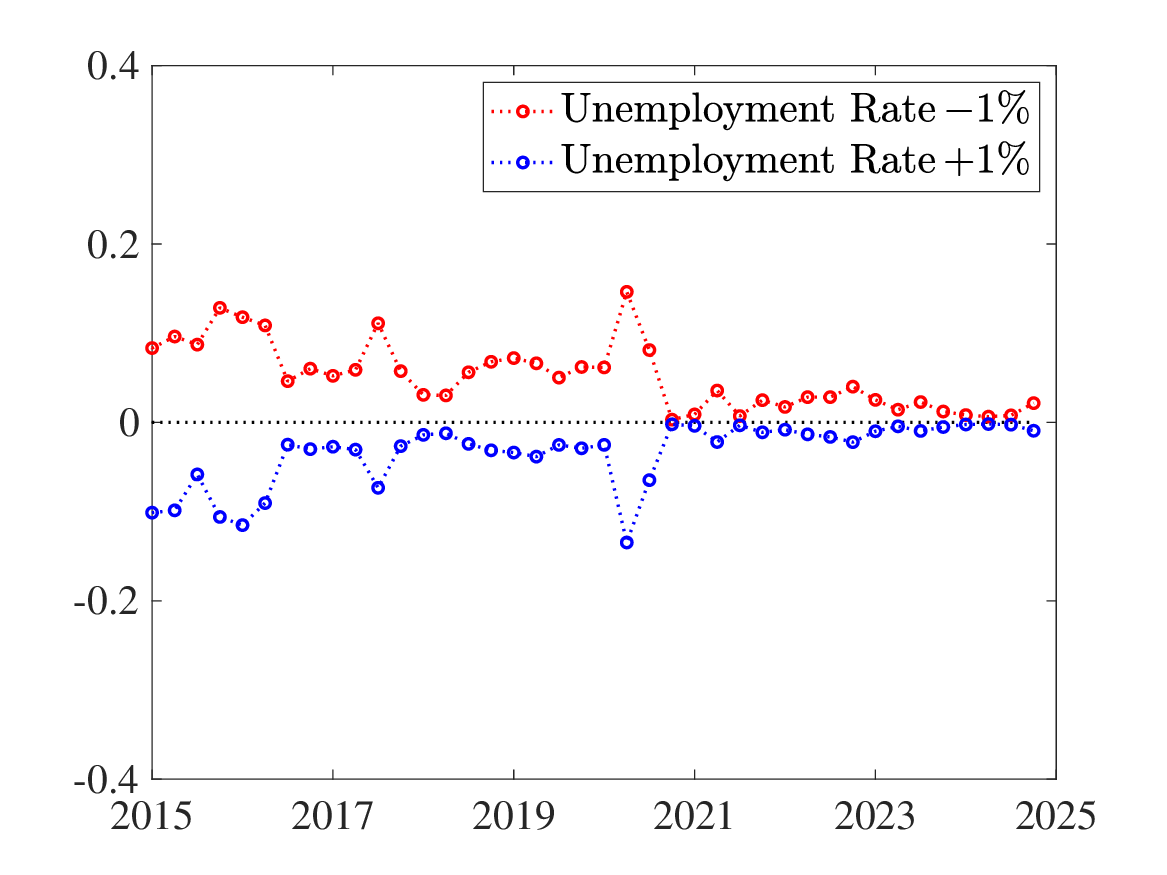}
     	\end{subfigure}
     	\hspace{-0.5cm}
     	\begin{subfigure}[b]{0.5\textwidth}
     		\centering
     		\caption{EIR: $P(\pi_{t+1}>4\%\mid x_t)$}	
     		\includegraphics[width=1\textwidth, height=0.7\textwidth]{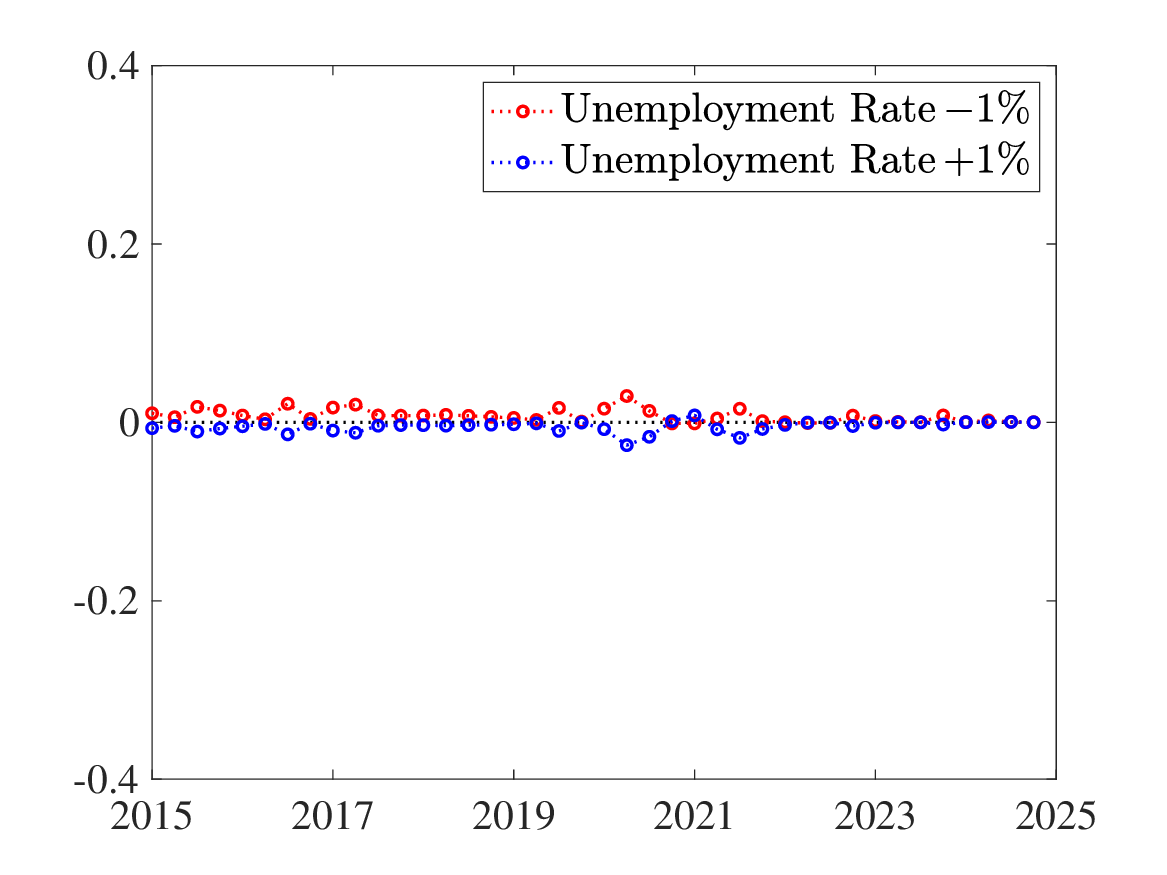}  
     	\end{subfigure}
     	\begin{minipage}{.9\linewidth} 
     		\linespread{1}\footnotesize
     		\textit{Notes}: This figure shows the change in deflation and excessive inflation risk probabilities resulting from a 1\% decrease (red) and a 1\% increase (blue) in unemployment rate, applied each quarter. Markers indicate the change in probability each quarter, with positive values representing increases and negative values representing decreases.
     	\end{minipage}
     	\label{fig: Counter-Unrate}
     \end{figure}

     Figure~\ref{fig: Counter-Unrate} presents the tail risk sensitivity to hypothetical $\pm1$ percentage-point shocks to the unemployment rate. The results demonstrate that changes in the unemployment rate exert negligible influence on excessive inflation risk throughout the sample, consistent with a very weak pass-through from labor market slack to the upper tail of the inflation distribution. For deflation risk, only small sensitivities are observable, primarily before 2021, with slightly larger (but still minor) effects visible during specific cyclical slowdowns in 2016 and the 2020 pandemic shock. The directional effects are broadly symmetric but consistently small in magnitude. These findings strongly complement the existing literature on the flattening of the PC by showing that the link between unemployment and inflation has significantly weakened even at the distributional extremes. The labor market appears to primarily affect the central tendency and the lower tail (deflation risk) only temporarily and weakly during exceptional economic shocks, while being largely irrelevant for upside inflation risk.

	\section{Conclusion}\label{sec: Con}
This paper advances a distributional perspective on inflation, showing that macroeconomic drivers shape the entire range of possible inflation outcomes rather than only the central tendency. Using the TVPDR framework, we capture the time-varying, nonlinear, and asymmetric nature of inflation dynamics.

Our empirical findings highlight the importance of distributional information for risk-management-oriented monetary policy. The framework tracks shifts in both deflation and high-inflation risks, particularly during periods of economic stress. By decomposing the contributions of underlying factors, we find that downside and upside inflation risks arise from distinct mechanisms: deflation risk is driven primarily by demand-side weakness and inflation persistence, whereas high-inflation risk is closely linked to supply-side disturbances, especially energy shocks. Moreover, the role of labor market slack is limited at the distributional extremes, indicating that the unemployment--inflation relationship weakens in the tails. By contrast, energy price inflation exhibits strongly asymmetric and time-varying effects on inflation risk, revealing nonlinear transmission channels that standard mean-based approaches may fail to capture. These findings imply that inflation stabilization requires monitoring not only expected inflation but also the evolving balance of downside and upside risks, since policy trade-offs and appropriate responses can differ substantially depending on whether inflation risks originate from demand-driven weakness or supply-driven disturbances.

	\clearpage
	\setstretch{1}
	\bibliographystyle{chicago}    
	\bibliography{TVP-DR-Inflation_ref}
	
    \newpage    
	\setstretch{1.3}
	\begin{center}
		\part*{Appendix}
	\end{center}
	
	\input{Appendix}


\end{document}

%% file: Appendix.tex
\setcounter{section}{0} 
\setcounter{figure}{0} 
\setcounter{table}{0} 
\setcounter{equation}{0} 
\setcounter{lemma}{0}\setcounter{page}{1}
\setcounter{proposition}{0} %
\renewcommand{\thepage}{A-\arabic{page}}
\renewcommand{\theequation}{A.\arabic{equation}}
\renewcommand{\thelemma}{A.\arabic{lemma}} 
\renewcommand{\theproposition}{A.\arabic{proposition}} 
\renewcommand\thesection{\Alph{section}} 
\renewcommand\thesubsection{A.\arabic{subsection}}
\renewcommand\thefigure{\thesection.\arabic{figure}}
\renewcommand\thetable{\thesection.\arabic{table}}

The appendices describe the estimation procedure and provide additional details on data and covariates. Specifically, Appendix \ref{sec: Estimation} describes the estimation procedure and algorithms, while Appendix \ref{sec: Appendix} provides details on the inflation measures and covariates used in the empirical analysis.

\section{Bayesian Estimation}\label{sec: Estimation}
This section presents a novel MCMC algorithm for estimating the TVPDR model by introducing a latent state-space model and a high-dimensional representation of the state-space model. More specifically, in subsection \ref{subsec: Representation}, we introduce the latent state-space model. Next, subsection \ref{subsec: MCMC} introduces the high-dimensional representation and describes our precision-based sampler. Finally, in subsection \ref{subsec: Monotonicity}, we show how to construct the entire conditional distribution using TVPDR and introduce an efficient algorithm that ensures the monotonicity condition on the conditional distribution function directly within the estimation process.

\subsection{A High-dimensional Representation}\label{subsec: Representation}
As discussed in Section \ref{sec: Model}, for an arbitrary location $y\in \mathcal{Y}_t$, model (\ref{TVPDR}) can be considered as a binary choice model with time-varying parameters for the binary outcome $\1\{Y_t\leq y\}$. A seminal paper \cite{albert1993bayesian} demonstrated an auxiliary variable approach for binary probit regression models that renders the conditional distributions of the model parameters equivalent to those under the Bayesian normal linear regression model with Gaussian noise. \cite{holmes2006bayesian} generalized the auxiliary variable approach to Bayesian logistic and multinomial regression models. \cite{polson2013bayesian} proposed a new data-augmentation strategy for fully Bayesian inference using Polya-Gamma latent variables that can be applied to any binomial likelihood parameterized by log odds like the logistic regression and negative binomial regression models.

Given the good properties of Gaussian distribution in the Bayesian framework, we develop our Bayesian inference for the TVPDR model with a focus on the probit link function setting. Following the auxiliary variable approach of \cite{albert1993bayesian}, we can study the proposed model via a latent Gaussian state-space model by assuming that there exists an unobserved continuous variable such that the binary event $\1\{Y_t\leq y\}$ occurs only if the latent variable exceeds a certain level. Specifically, we can consider a latent variable $Y_{y, t}^{*}$ for Model (\ref{TVPDR}) that satisfies $\1 \{Y_{t} \leq y\}=\1 \{Y_{y, t}^{*}\geq 0\}$, and defined by the following Gaussian state-space model,

\begin{equation}
	\label{eq:  Latent-Model}
	\begin{aligned}
		Y_{y, t}^{*} &= g(X_{t})^\top\beta_{y, t} + \varepsilon_{y, t}, & \varepsilon_{y, t} &\sim \mathcal{N}(0,1), \\
		\beta_{y, t} &= \beta_{y,t-1} + \eta_{y, t}, & \eta_{y, t} &\sim \mathcal{N}(\mathbf{0}_{d},\Sigma_{y}).
	\end{aligned}
\end{equation}

Equivalently, given the observed data and parameters, the latent $Y_{y, t}^{*}$ has the following conditional distributions,
\begin{align}\label{eq: latent}
	Y_{y, t}^{*}\mid Y_t, X_t, \beta_{y,t}\sim\begin{cases}
		\mathcal{N}_{[0,\infty)}\big(g(X_{t})^\top\beta_{y, t},1\big), & \text{if}\ \ \1 \{Y_{t}\leq y\} \\
		\mathcal{N}_{(-\infty,0)}\big(g(X_{t})^\top\beta_{y, t},1\big), & \text{if}\ \ \1 \{Y_{t}>y \},
	\end{cases}
\end{align}
where $\mathcal{N}_{A}$ denotes a truncated normal distribution on set $A$. 

Based on this framework, we introduce a precision-based MCMC algorithm to estimate all model parameters efficiently. It is worth noting that while our algorithm is developed with a focus on the probit-link case, it can be easily generalized to the logit-link case using ideas introduced by \cite{holmes2006bayesian} and \cite{polson2013bayesian}.

\subsection{The precision-based sampler for TVPDR}\label{subsec: MCMC}
Assume that we have observations of $\{Y_t, X_{t}\}$ in periods $t=1,\ldots, T$ available for estimating the unknown parameters. Given the simulated latent variables $Y_{y, t}^{*}$, our model becomes a linear Gaussian state-space model. For which, the standard approach in Bayesian literature for sampling the unobserved time-varying parameters is to use Kalman filtering-based algorithms \citep{carter1994gibbs,durbin2002simple}. There are recent advances in the MCMC literature that leverage the relatively sparse precision matrix to gain substantial computational advantages \citep{chan2009efficient,chan2023conditional}. To utilize such a precision-based sampler, we rewrite the latent state-space model as a high-dimensional static regression with more covariates than observations by stacking all $T$ observations together.

Specifically, let $\boldsymbol{\beta}_{y}=(\beta_{y,1}^\top,\beta_{y,2}^\top,\ldots,\beta_{y,T}^\top)^\top\in\mathbb{R}^{Td}$ and $\boldsymbol{\eta}_{y} =(\eta_{y,1}^\top,\eta_{y,2}^\top,\ldots,\eta_{y,T}^\top)^\top\in\mathbb{R}^{Td}$, it follows from the random walk assumption in (\ref{eq: TVP}) that
\begin{align*}
	\mathbf{H}\boldsymbol{\beta}_{y}=\boldsymbol{\eta}_{y}\sim \mathcal{N}\big(\mathbf{0}_{Td}, \boldsymbol{\Omega}_{y}\big),
\end{align*}
where 
$
\mathbf{H}=(\mathbb{I}_T-\mathbb{I}_{T,-1}) \otimes \mathbb{I}_{d}
$
and $\boldsymbol{\Omega}_{y}=\mathbb{I}_{T}\otimes \Sigma_{y}$. Note that both $\mathbf{H}$ and $\boldsymbol{\Omega}_{y}$ are $Td\times Td$ banded matrices\footnote{Banded matrix refers to a sparse matrix whose non-zero elements are arranged along a diagonal band.}.
Furthermore, stacking $\mathbf{Y}_{y}^{*}=\big(Y_{y,1}^{*},Y_{y,2}^{*},\ldots,Y_{y,T}^{*}\big)^\top\in\mathbb{R}^{T},
$
the Gaussian state-space model (\ref{eq: Latent-Model}) can be written as 
\begin{align}\label{eq: Latent Model All}
	&\mathbf{Y}_{y}^{*} =\mathbf{X}\boldsymbol{\beta}_{y}+\boldsymbol{\varepsilon}_{y},\ \ \boldsymbol{\varepsilon}_{y}\sim \mathcal{N}(\boldsymbol{0},\mathbb{I}_{T}),\\
	&\boldsymbol{\beta}_{y} \mid \boldsymbol{\Omega}_{y}\sim \mathcal{N}\big(\mathbf{0}_{d},(\mathbf{H}^\top\boldsymbol{\Omega}_{y}^{-1}\mathbf{H})^{-1}\big)\label{Beta},
\end{align}
where
$
\mathbf{X}=\diag (g(X_{1})^\top, \ldots, g(X_{T})^\top)
$
is a banded matrix of dimension $T\times Td$, 
and $\boldsymbol{\varepsilon}_{y}=(\varepsilon_{y,1},\varepsilon_{y,2},...,\varepsilon_{y,T})^\top\in\mathbb{R}^{T}$.

This high-dimensional representation of the latent Gaussian state-space model allows us to develop an efficient precision-based MCMC algorithm, substantially speeding up computations. First, we can consider (\ref{Beta}) as a prior for $\boldsymbol{\beta}_{y}$. Since the distribution of the latent $\mathbf{Y}_{y}^*$ conditional on $\boldsymbol{\beta}_{y}$ is Gaussian, a simple application of Bayes' theorem
implies that the conditional posterior distribution of $\boldsymbol{\beta}_{y}$ is also Gaussian
\begin{align}\label{eq: post-beta}
	\boldsymbol{\beta}_{y}\mid \mathbf{X},\mathbf{Y}_{y}^*, \boldsymbol{\Omega}_{y}  \sim \mathcal{N}\big(\mathbf{\boldsymbol{\mu}}_{y},\mathbf{K}_{y}^{-1}\big),
\end{align}
where 
\begin{align}
	\boldsymbol{\mu}_{y} =\mathbf{K}_{y}^{-1}\left(\mathbf{X}^\top\mathbf{Y}_{y}^{*}\right),\ \ \ \ \mathbf{K}_{y} =\mathbf{X}^\top\mathbf{X}+\mathbf{H}^\top\boldsymbol{\Omega}_{y}^{-1}\mathbf{H}.
\end{align}
Given that $\mathbf{X}, \mathbf{H}$ and $\boldsymbol{\Omega}_{y}$ are all banded matrices, the precision matrix $\mathbf{K}_{y}$ is also banded. Therefore, given the draws of the latent variables, we can use the precision-based sampler of \cite{chan2009efficient} to draw the time-varying parameters $\boldsymbol{\beta}_{y}$ efficiently. 

In order to regularize the degree of time variation of the parameters, the TVP models are typically equipped with tightly parameterized prior distributions for $\Sigma_{y}$ that favor gradual changes in the parameters \citep{nakajima2011time,primiceri2005time}. One conventional candidate is the inverse Gamma prior, where the covariance matrix is assumed to be diagonal, that is, $\Sigma_{y}=\diag\big(\sigma_{y,1}^{2},\sigma_{y,2}^{2},\ldots,\sigma_{y,d}^{2}\big)$. For each $\sigma_{y,i}^2, i=1,\ldots, d$, we use independent weakly informative inverse Gamma priors $\sigma_{y,i}^{2}\sim\mathcal{IG}(\nu_{y,i},S_{y,i})$, where $\nu_{y,i}$ is the degree of freedom parameter and $S_{y,i}$ is the scale parameter. The posterior distributions are given by
\begin{align}\label{eq: post-sigma}
	\sigma_{y,i}^{2}\mid\boldsymbol{\beta}_{y}
	\sim\mathcal{IG}\left(\nu_{y,i}+\frac{T-1}{2},S_{y,i}+\frac{1}{2}\sum_{t=2}^{T}(\beta_{y,t,i}-\beta_{y,t-1,i})^{2}\right).
\end{align}
The procedures for deriving the posterior distributions (\ref{eq: post-beta}) and (\ref{eq: post-sigma}) are standard and can be found in \cite{koop2003bayesian}.

\subsection{Monotonicity of the Distribution Function}\label{subsec: Monotonicity}
Traditionally, one could apply the proposed model and MCMC algorithm to estimate the conditional distribution function $F_{Y_t|X_t}(\cdot|X_{t})$ on a sequence of fine enough discrete points over the support $\mathcal{Y}_t$. The collection of estimation results can approximate the entire conditional distribution of $Y_t$.

One important property that characterizes $F_{Y_t|X_t}(\cdot|X_{t})$ is monotonicity, i.e., the conditional distribution function is non-decreasing by definition. Yet, the distribution functions obtained by estimating $F_{Y_t|X_t}(y_j|X_{t})$ for each $y_j\in\mathcal{Y}_t, j=1,\ldots,K$, independently do not necessarily satisfy monotonicity in finite samples. The standard strategy used in DR literature monotonizes the conditional distribution values at different locations using a rearrangement method proposed by \cite{chernozhukov2009improving}. In the Bayesian context, a naive way to ensure monotonicity is to first run MCMC estimation for each discrete point $y_j$ independently. For $X_t=x_t$, in each iteration, we can evaluate $F_{Y_t|X_t}(y_j|x_t)$ for $j=1,\ldots,K$ using the draws of $\beta_{y_j, t}$, and rearrange these distribution values using the two-step approach. However, this method has limitations when $\beta_{y_j, t}$ for different $y_j$ have varying convergence rates. To address this challenge, we introduce a MCMC algorithm that estimates all time-varying parameters across different locations simultaneously while explicitly imposing a monotonicity condition on the conditional distribution function.

Under the TVPDR model (\ref{TVPDR}), since the link function $\Lambda$ is an non-decreasing transformation, the monotonicity of the conditional distribution at $y_j\in\mathcal{Y}_t, j=1,\ldots, K$, can be ensured by imposing the following constraint:
\begin{align}
	\label{eq: monotone}
	\mathbf{X}\boldsymbol{\beta}_{y_{1}} \preceq \ldots \preceq \mathbf{X}\boldsymbol{\beta}_{y_K}.
\end{align}
Let
$
\boldsymbol{\beta}=(\boldsymbol{\beta}_{y_1},....,\boldsymbol{\beta}_{y_K})\in \mathbb{R}^{Td\times K},
$
the constraint can be equivalently expressed as the following set
\[
\mathcal{S} \doteq \left\{ \boldsymbol{\beta}\in \mathbb{R}^{Td\times K}:  \left(M \otimes \mathbf{X}\right) \vecop( \boldsymbol{\beta})\succeq \mathbf{0}_{T(K-1)} \right\},
\]
where $M$ is a selecting matrix defined by
$
M=[\mathbf{0}_{K-1},\mathbb{I}_{K-1}]-[\mathbb{I}_{K-1},\mathbf{0}_{K-1}].
$
Thus, if one is to naively sample them jointly, we are facing the following conditional posterior
\[
\boldsymbol{\beta}|\mathbf{X},\mathbf{Y}^*_{y_1},...,\mathbf{Y}^*_{y_K},  \boldsymbol{\Omega}_{y_1}..., \boldsymbol{\Omega}_{y_K}\sim \mathcal{N}_{\mathcal{S}}\left(
\begin{bmatrix}
	\boldsymbol{\mu}_{y_1}\\
	\boldsymbol{\mu}_{y_2}\\
	...\\
	\boldsymbol{\mu}_{y_K}
\end{bmatrix},  \begin{bmatrix}
	\mathbf{K}^{-1}_{y_1}&\mathbb{O}_{Td}&...&\mathbb{O}_{Td}\\
	\mathbb{O}_{Td} &\mathbf{K}^{-1}_{y_2}&...&\mathbb{O}_{Td}\\
	...&...&...&...\\
	\mathbb{O}_{Td}&\mathbb{O}_{Td}&...&\mathbf{K}^{-1}_{y_K}
\end{bmatrix}
\right).	
\]
Given the presence of a total of $T(K-1)$ constraints, the approach becomes unviable when dealing with DR models featuring constant parameters. In such cases, the number of unknown parameters ($Kd$) is substantially smaller than the total number of imposed constraints. However, by allowing parameters to vary over time, we gain the ability to sample all $\boldsymbol{\beta}_{y}$ simultaneously with the $T(K-1)$ constraints, when constructing the complete conditional distribution. It is important to highlight that, conceptually, this strategy remains effective as long as at least one of the parameters is assumed to be time-varying. For instance, it is adequate to introduce time variation only in the intercept parameter, especially when the research goal is to capture the dynamic changes in the conditional distribution of a time series.

Compared to the standard strategy that monotonizes the conditional distribution values at different locations using the rearrangement method, this new method allows us to sample $\boldsymbol{\beta}$ by sampling $\boldsymbol{\beta}_{y_j}$ from $j=1$ to $j=K$ sequentially from their posterior distributions
\[
\boldsymbol{\beta}_{y_j}\mid \mathbf{X},\mathbf{Y}_{y_j}^*, \boldsymbol{\Omega}_{y_j}  \sim \mathcal{N}\big(\mathbf{\boldsymbol{\mu}}_{y_j},\mathbf{K}^{-1}_{y_j}\big),
\]
subject to the following constraint
\begin{align}\label{eq: con1}
	\mathbf{X}\boldsymbol{\beta}_{y_{j-1}} \preceq \mathbf{X}\boldsymbol{\beta}_{y_{j}} \preceq \mathbf{X}\boldsymbol{\beta}_{y_{j+1}}.
\end{align}	
In practice, achieving this involves sampling $\boldsymbol{\beta}_{y_j}$ from a $Td$-dimensional truncated Gaussian distribution, which is quite computationally challenging given the relatively high dimension. Here, we introduce a strategy that makes the simulation feasible and efficient by exploiting a special structure of our constraint.

Without loss of generality, we assume that the intercept term is considered in the model, that is,  $g(X_t)$ includes 1 as the first element. We first separate all intercept parameters and the other parameters of $\boldsymbol{\beta}_{y_j}$ into the following two vectors,
\[
\boldsymbol{\beta}_{y_j}^{(1)} := M_1 \boldsymbol{\beta}_{y_j},\ \ \ \ \boldsymbol{\beta}_{y_j}^{(2)} := M_2 \boldsymbol{\beta}_{y_j},
\]
where
$
M_1 = \mathbb{I}_{T}\otimes [1, \textbf{0}_{d-1}]
$
and $
\ M_2 = \mathbb{I}_{T}\otimes [\textbf{0}_{d-1}, \mathbb{I}_{d-1}]
$
are selection matrices that select the intercepts and the coefficients other than the intercepts, respectively. 
The constraint in (\ref{eq: con1}) can be rewritten as linear constraints imposed on all intercept parameters of $\boldsymbol{\beta}_{y_j}$, as described by the following set

\[
\mathcal{S}_j \doteq \left\{ \boldsymbol{\beta}_{y_j}^{(1)}\in \mathbb{R}^{T}:  \mathbf{X}\boldsymbol{\beta}_{y_{j-1}} - \mathbf{X} M_2 ^\top\boldsymbol{\beta}_{y_j}^{(2)}  \preceq \boldsymbol{\beta}_{y_j}^{(1)} \preceq  \mathbf{X}\boldsymbol{\beta}_{y_{j+1}} - \mathbf{X} M_2^\top\boldsymbol{\beta}_{y_j}^{(2)} \right\}.
\]

This enables us to sample $\boldsymbol{\beta}_{y_j}$ via an efficient two-step sampling approach that greatly reduces the dimension of the truncated Gaussian distribution required in the simulation. More specifically, based on the marginal-conditional decomposition of $\boldsymbol{\beta}_{y_j}$, we can first sample $\boldsymbol{\beta}_{y_j}^{(2)}$ from its unconstrained marginal distribution using the precision-based sampler of \cite{chan2009efficient}. Conditional on simulated $\boldsymbol{\beta}_{y_j}^{(2)}$, we can then sample $\boldsymbol{\beta}_{y_j}^{(1)}$ from a $T$-dimensional truncated Gaussian distribution, where methods like the minimax tilting method of \cite{botev2017normal} can be directly applied. The algorithm for sampling $\boldsymbol{\beta}$ simultaneously with a monotonicity constraint is described in Algorithm \ref{alg2}.

The proposed algorithm for ensuring monotonicity is applicable across a diverse spectrum of regression models to estimate monotonic functions, as described in (\ref{eq: monotone}). This application assumes that the coefficients $\boldsymbol{\beta}$ are conditionally Gaussian. An example of a regression model that falls within this structure is Bayesian QR \citep{korobilis2021time}. Quantile functions exhibit monotonic behavior as the quantile parameter, typically ranging from 0 to 1, increases. There has been a proliferation in the application of the Gibbs sampling algorithm for Bayesian 	QR, which is based on augmenting the Asymmetric Laplace density within a conditionally Gaussian structure \citep{kozumi2011gibbs}. This proliferation has placed a specific emphasis on focusing on one quantile level at a time. However, it is worth noting that, to date, we have not encountered any MCMC algorithms for ensuring monotonicity when estimating multiple quantiles, as maintaining order across MCMC samplers in such cases is not a straightforward task. Algorithm \ref{alg: monotone} provides a solution for addressing this class of problems.

\begin{algorithm}[H]
	\caption{Sampling $\boldsymbol{\beta}$ simultaneously with monotonicity constraint}
	\label{alg2}
	In each iteration, from $j=1$ to $j=K$, \\	 
	\textbf{Step 1}.  Sample $\boldsymbol{\beta}_{y_j}^{(2)} $ from its unconstrained marginal posterior distribution
	\[
	\boldsymbol{\beta}_{y_j}^{(2)}\mid \mathbf{X},\mathbf{Y}_{y_j}^*, \boldsymbol{\Omega}_{y_j}  \sim \mathcal{N}\big(\mathbf{\boldsymbol{\mu}}_{y_j}^{(2)},\mathbf{K}_{y_j}^{(2)-1}\big),
	\]
	where
	\begin{align*}
		\mathbf{\boldsymbol{\mu}}_{y_j}^{(2)}=M_2 \mathbf{\boldsymbol{\mu}}_{y_j},\ \ \ \ \mathbf{K}_{y_j}^{(2)}=	M_2 \mathbf{K}_{y_j} M_2^\top.
	\end{align*}
	
	\textbf{Step 2}. Sample $\boldsymbol{\beta}_{y_j}^{(1)}$ from its constrained conditional posterior distribution
	\[
	\boldsymbol{\beta}_{y_j}^{(1)}\mid \boldsymbol{\beta}_{y_j}^{(2)},  \boldsymbol{\beta}_{y_{j-1}},  \boldsymbol{\beta}_{y_{j+1}}, \mathbf{X},\mathbf{Y}_{y_j}^*, \boldsymbol{\Omega}_{y_j}  \sim \mathcal{N}_{\mathcal{S}_j}\big(\mathbf{\boldsymbol{\mu}}_{y_j}^{(1)},\mathbf{K}_{y_j}^{(1)-1}\big),
	\]
	where
	\begin{align*}
		& \mathbf{K}_{y_j}^{(1)} = M_1\mathbf{K}_{y_j}M_1^\top,\\
		& \mathbf{\boldsymbol{\mu}}^{(1)}_{y_j} =  \mathbf{K}_{y_j}^{(1)-1}M_1\mathbf{K}_{y_j}\big(\boldsymbol{\mu}_{y_j}-M_2^\top\boldsymbol{\beta}^{(2)}_{y_j}\big).
	\end{align*}
	\label{alg: monotone}
\end{algorithm}

\renewcommand{\thepage}{B-\arabic{page}}
\renewcommand{\theequation}{B.\arabic{equation}}
\renewcommand{\thelemma}{B.\arabic{lemma}} 
\renewcommand{\theproposition}{B.\arabic{proposition}} 
\renewcommand\thesection{\Alph{section}} 
\renewcommand\thesubsection{B.\arabic{subsection}}
\renewcommand\thefigure{\thesection.\arabic{figure}}
\renewcommand\thetable{\thesection.\arabic{table}}

\section{Data Description}\label{sec: Appendix}
This appendix provides details on the inflation measures and covariates used in the empirical analysis. All data are obtained from the Federal Reserve Bank of St. Louis’s FRED database.

\subsection{Inflation}
\begin{figure}[H]
	\captionsetup[subfigure]{aboveskip=-2pt,belowskip=0pt}
	\centering
	\caption{Inflation Realizations}
	\begin{subfigure}[b]{0.5\textwidth}
		\centering
		\caption{Headline Inflation}				 
		\includegraphics[width=1\textwidth, height=0.7\textwidth]{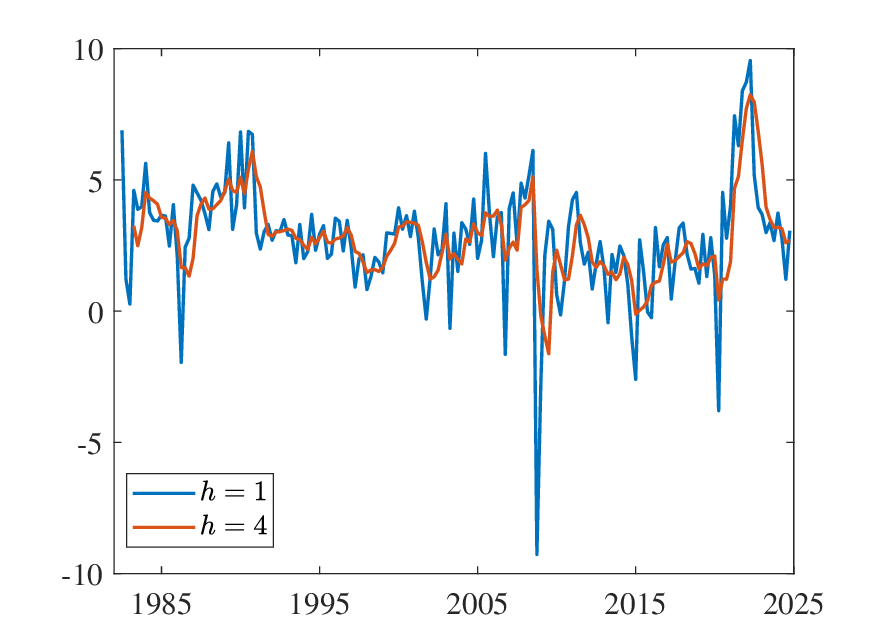}
	\end{subfigure}
	\hspace{-0.5cm}
	\begin{subfigure}[b]{0.5\textwidth}
		\centering
		\caption{Core Inflation}	
		\includegraphics[width=1\textwidth, height=0.7\textwidth]{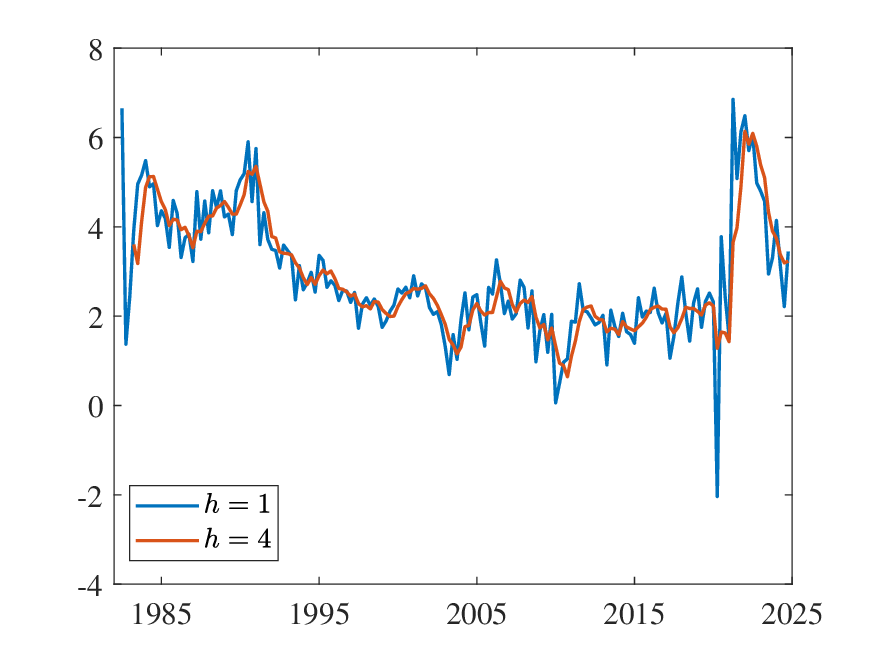} 
	\end{subfigure}
	\begin{minipage}{0.9\linewidth}
		\linespread{1}\footnotesize
		\textit{Notes}: $\pi_{t+h}=(400/h) \ln (P_{t+h}/P_t)$
	\end{minipage}
	\label{fig: Inflation}
\end{figure}

We consider both headline and core inflation, constructed from the following two Consumer Price Index (CPI) series:
\begin{itemize}
	\item \textbf{Consumer Price Index for Headline Inflation} ($P_t$): Consumer Price Index for All Urban Consumers: All Items in U.S. City Average, Index 1982-1984=100, Seasonally Adjusted. Source:  FRED, `CPIAUCSL'.
	\item \textbf{Consumer Price Index for Core Inflation} ($P_t$): Consumer Price Index for all urban consumers: all items less food and energy (1982-84=100). Source:  FRED, `CPILFESL'.
\end{itemize} 
Figure \ref{fig: Inflation} plots the time series of headline and core inflation defined as $\pi_{t+h}=(400/h) \ln (P_{t+h}/P_t)$ for horizons $h=1, 4$.

\subsection{Covariates}
Table~\ref{Tab: Covariate} lists all covariates used in the empirical analysis, along with their descriptions and data sources.
\begin{table}[H]
	\footnotesize
	\caption{Covariates}
	\centering
	\begin{tabular}{lll}
		\hline \hline
		Covariates             & Description                                                                  & FRED Source \\ \hline
		Inflation Expectation  & 5-year expected inflation, Percent           & `T5YIFR'    \\
		Unemployment Rate      & Civilian Unemployment Rate, Percent                  & `UNRATE'    \\
		Energy Price Inflation & Consumer Price Index for All Urban Consumers: Energy in U.S.                 & `CPIUFDSL'  \\
		& City Average, Continuously Compounded Annual Rate of Change  &             \\
		Food Price Inflation   & Consumer Price Index for All Urban Consumers: Food in U.S.                   & `CPIUFDSL'  \\
		& City Average, Continuously Compounded Annual Rate of Change &             \\
		Financial Condition    & Chicago Fed National Financial Conditions Index    & `NFCI'      \\
		Federal Funds Rate     & Federal Funds Effective Rate, Percent      & `FEDFUNDS'  \\
		Real GDP Growth        & Real Gross Domestic Product, Percent Change, Annual Rate & `GDPC1'     \\ \hline
	\end{tabular}       	
	\label{Tab: Covariate}       	
\end{table}